\documentclass[final,3p,times,12pt]{elsarticle}

\usepackage{amsmath} 

\usepackage{amssymb}
\usepackage{multirow}
\usepackage[table,xcdraw]{xcolor}

\usepackage{subfig}
\usepackage{graphicx}
\usepackage{stfloats}
\usepackage{fixltx2e}

\usepackage{lineno}

\usepackage{subcaption}
\usepackage{graphicx}

\journal{}
\begin{document}

\begin{frontmatter}



\title{The Role of Net-Zero Energy Residential Buildings in Florida's Energy Transition: Economic Analysis and Technical Benefits}



\author[]{Hamed Haggi, James M. Fenton}

\affiliation{organization={Florida Solar Energy Center, University of Central Florida},
            city={Cocoa},
            state={Florida},
            country={USA}}
            
\ead{hamed@ucf.edu, jfenton@fsec.ucf.edu}

\begin{abstract}

Net-zero emission policies, coupled with declining costs of renewables, battery storage, and electric vehicles, require both direct and indirect electrification (such as green hydrogen) of the energy infrastructure to address climate change impacts, improve resilience, and lower energy expenses. In this paper, a comprehensive techno-economic analysis was performed that integrated net-zero energy buildings within the context of Florida’s energy transition by the year 2050. The analysis compares the monthly cost savings of owning photovoltaic (PV) on the roof, battery storage, and Electric Vehicle (EV) to the cost of buying electricity from the grid for both existing and newly built Orlando homes. The impact of income tax credits (ITC) and energy efficiency improvements on monthly savings was taken into account. The levelized cost of electricity for residential PV and PV + battery systems was determined and extended to provide a cost equivalent to gasoline for an average EV. Rooftop solar fueling an EV for 10,000 miles saves \$100 per month over purchasing gasoline.  Today, Florida residents can save on their monthly costs of electricity and gasoline if they have both PV on their roof and a battery system (sized for average daily residence load), and benefit from the federal ITC.  Florida residents can cost-effectively transition to their own solar, battery storage, and electric vehicle to home systems. Allowing utility-scale solar to be used for on-site EV fast-charging and hydrogen production through electrolysis for fuel cell vehicles or blending with natural gas, minimizing the need for electricity transmission and distribution costs across the power grids. 

\end{abstract}









\end{frontmatter}

\section{Introduction}

The global energy system currently accounts for approximately 75\% of greenhouse gas emissions, making it a pivotal factor in addressing society's greatest challenge: averting the worst effects of climate change \cite{bouckaert2021net}. Achieving Net Zero Emissions (NZE) by 2050, which is essential to limit the long-term increase in average global temperatures to 1.5°C \cite{anika2022prospects} \cite{haggi2023p2p}, necessitates a comprehensive transformation in how to produce, transport, and consume energy in both the United States and the state of Florida. This transformation calls for a concerted effort to advance clean energy technologies throughout the decade leading up to 2030. While there has been promising growth in renewable energy over the past two decades moving towards a decarbonized society, significant challenges persist in achieving substantial penetration into the grid-based energy market. Despite improvements in Florida's building energy efficiency through the adoption of more stringent energy codes, the building sector continues to be a significant source of greenhouse gas emissions. These challenges encompass: 1) Ensuring resiliency and security as solar capacity is integrated into the grid;
2) Establishing the value proposition of energy storage;
3) Ensuring highly reliable and secure interconnection of solar energy resources;
4) Optimizing the performance and durability of solar and energy systems;
5) Harnessing the benefits of vehicle electrification for grid interactions;
6) Enhancing building energy performance;
7) Facilitating the introduction of technological innovations to the market;
and 8) Developing a skilled workforce for the energy industry.

\begin{figure}
\centering
\footnotesize
	\includegraphics[width=5in]{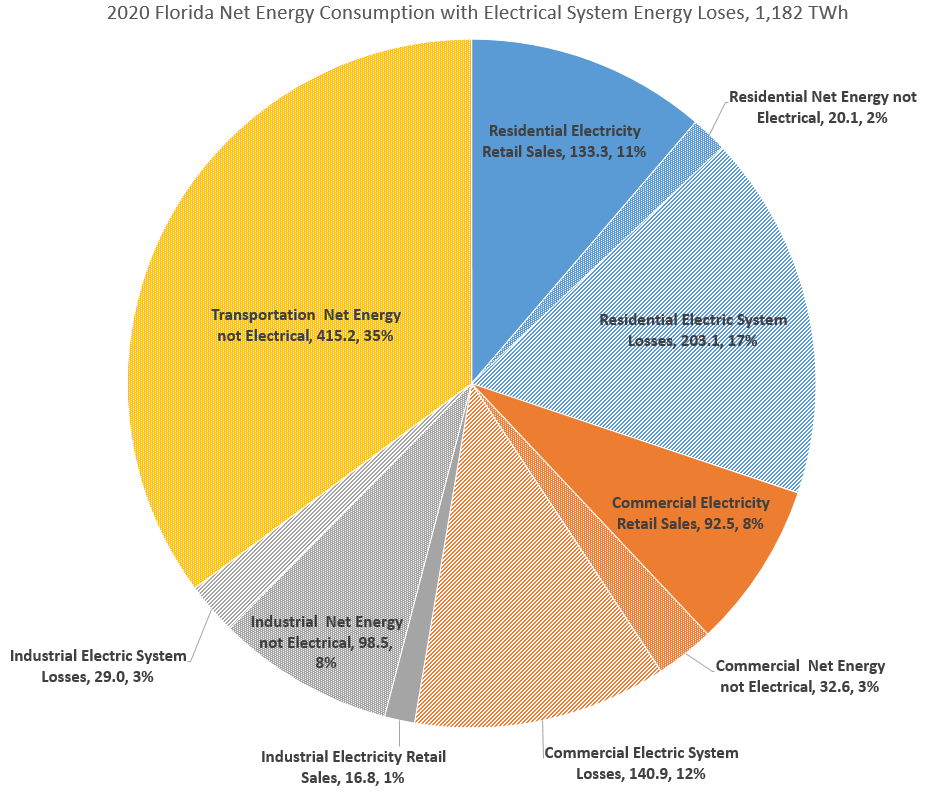}
	\caption{Florida net energy consumption considering electrical system losses.}
    \label{Piechart_Fig1}
\end{figure}

In 2020, Florida's net end-use energy consumption across the Residential, Commercial, Industrial, and Transportation sectors totaled 809 TWh, with 242 TWh attributed to electricity retail sales. The overall energy losses within the electrical systems of these sectors amounted to 373 TWh, resulting in a combined end-use energy consumption of 1,182 TWh \cite{EIAFlorida}. Figure \ref{Piechart_Fig1} shows the breakdown of Florida's electricity retail sales, fossil fuel energy usage, and electric system losses for each of these sectors, both in TWh and as a percentage of the total end-use energy consumption. Notably, the efficiency of electric utility delivery to these sectors was only 40\%, calculated as 242 / (242 + 373) in TWh. In the transportation sector, fossil fuel consumption reached 415.2 TWh, with only 0.1 TWh attributed to electric consumption. A mere 21\% was effectively utilized for vehicle mileage. Unfortunately, Florida's Residential 203.1 TWh of electric systems losses exceeded the 153.4 TWh of residential sector electricity consumption.
Replacing this 153.4 TWh of residential electricity with on-site solar energy generation and battery energy storage, Florida preserves the current residential electricity usage, has reliable backup power, and also eliminates the 364.5 TW of residential electric system losses. 

\begin{figure}
\centering
\footnotesize
	\includegraphics[width=6.5in,height= 3in]{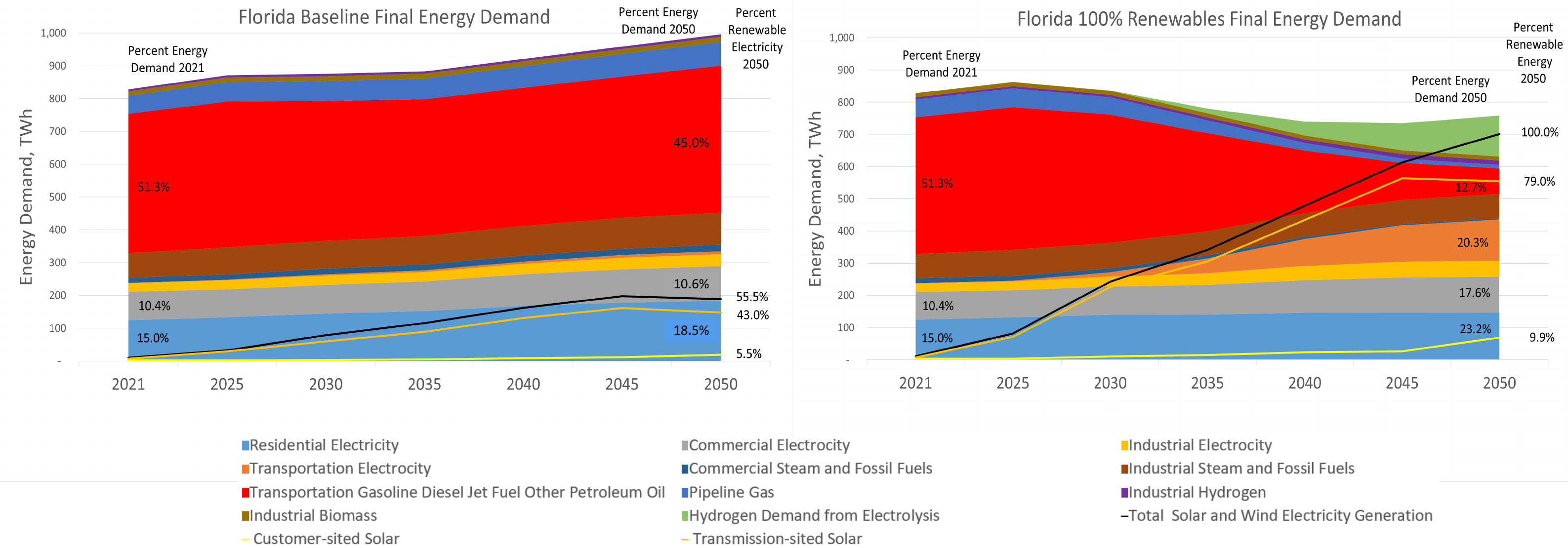}
	\caption{Florida baseline final energy demand vs. 100\% renewable scenario.}
    \label{FLA_100_baseUpdate}
\end{figure} 

There have been several global studies that have outlined pathways to achieve net-zero carbon dioxide emissions or 100\% renewable energy by 2050. However, many of these studies lack the level of detail, necessary for infrastructure and policy planning at the state level. Jacobson et al. in \cite{jacobson2022zero} explored the grid stability in the US after transitioning all sectors, including electricity production, transportation, buildings, and industrial energy, to 100\% renewable electricity. The authors also provided data specific to Florida as supplementary material for their paper. The main results for transitioning Florida to 100\% Renewables over business as usual were found to be: 326 TWh of utility photovoltaic (PV), 115 TWh of rooftop PV, 100\% grid stability at low cost; creation of 393,000 more long-term, full-time jobs than lost; saving of 2,840 lives from air pollution per year in 2050; elimination of 283 million tons-CO2 per year in 2050; reduction in 2050 all-purpose, end-use energy requirements by 52.8\% (430 TWh instead of 908 TWh); reduction in 2050 annual energy costs by 55.6\% (from \$102.4 to \$45.4 B/y); reduction in annual energy, health, plus climate costs by 84.7\% from \$298 to \$45.4 B/y; while requiring 1.37\%  of Florida land for footprint, and 0.38\% for spacing. Williams et al. in \cite{williams2021carbon} outlined eight detailed scenarios for the United States to achieve net-zero carbon dioxide emissions, and they also provide data specific to Florida. Their findings indicate that achieving NZE will come at a net cost of less than 1.2\% of GDP, not accounting for climate benefits, with over 80\% of the energy coming from wind and solar sources. Two of these scenarios, the Baseline and the 100\% Renewables scenario with hydrogen generation from electricity, were used to generate the figures for Florida's Final Energy Demand as shown in Figure \ref{FLA_100_baseUpdate}. In Figure \ref{FLA_100_baseUpdate}, the total energy demand for the year 2021 was 828 TWh, of which 234 TWh was electricity. In the Baseline scenario, the projected energy demand for 2050 is 996 TWh, representing a 20\% increase from 2021. However, in the 100\% Renewables scenario, the 2050 energy demand is estimated at 759 TWh, with 128 TWh allocated for Hydrogen Production by Electrolysis, marking an 8.3\% decrease from 2021. In 2021, solar energy accounted for only 4.8\% of electricity demand (11.2 TWh), which is equivalent to just 1.4\% of the total energy demand. In the Baseline scenario, solar and wind together makeup 55.5\% of the 2050 electricity demand. In the 100\% Renewables scenario, 92.5\% of the 2050 energy demand (701 TWh) is expected to come from solar and wind sources. Out of this, 69.3 TWh is attributed to customer-sited solar, indicating that Florida utilities will be producing 632 TWh of solar and wind energy, which is 2.7 times the amount of electricity they produce in 2021. By 2050, residential electricity demand is projected to account for only 19.3\% of the total energy demand in the 100\% Renewables Scenario. Even if all building electricity demands were met by customer-owned solar generation (225.5 TWh), utilities would still generate at least 476 TWh of solar and wind energy (possibly more to account for transmission and distribution losses), which is twice the amount of electricity the utilities produce in Florida in 2021.

\vspace{2mm}

There are several approaches to minimize the electricity demand of residential buildings such as employing energy-conserving architectural designs, adopting high-performance appliances, optimizing heating, ventilation, and air conditioning (HVAC) systems, cultivating energy-aware resident conduct, and integrating intelligent control mechanisms \cite{wu2018selecting}. Energy-efficient architectural design for new construction or retrofit of homes augments the structural envelopes of buildings by increasing thermal insulation, advanced window glazing, and the assimilation of reflective or eco-friendly roofing technologies, all intended to abate the impact of solar heat gain \cite{li2019energetic}. Research studies on Net Zero Energy Buildings (NZEBs) with PV and battery systems typically focus on single-day operations and use financial metrics like net present value (NPV) to assess their feasibility or consider hybrid systems with considering electric vehicles for resilience improvement in long-duration outages \cite{gorman2023county}. For example, Kim et al. \cite{kim2023economic} explored the economic feasibility of implementing NZEBs in US homes using a heat pump system alongside federal government-supported solar and geothermal technologies. They highlighted the challenges of achieving "net-zero emission" and analyzed the payback periods of various NZEB scenarios, considering future changes in technology and policy needed to reach the net-zero emission target by 2050. Sohani et al. \cite{sohani2023techno} explored the benefits of PV and building integrated PV thermal hybrid systems for electricity generation, with excess electricity used to drive a hot and cold water storage system for a residential building in the city of Tehran. A techno-economic analysis was conducted by \cite{KLEINEBRAHM2023} to determine the most cost-effective PV solution for achieving a self-sufficient energy supply in buildings across Central Europe. The simulation results indicate that 53\% of the 41 million single-family residences have the potential to operate independently from the power grid and projections suggest that by 2050, this figure could rise to 75\%. Cucchiella et al. \cite{cucchiella2018solar} investigated the economic viability of residential buildings equipped with 3-kW PV and lead-acid battery storage systems in Italian homes. Zakeri et al. \cite{zakeri2021policy} explored different policy incentives for energy efficiency improvements in U.K. residential buildings. Their study aimed to minimize electricity consumption by 84\% using a PV and battery storage system when optimized. Hale et al. \cite{hale2018integrating} focused on the future challenges of Florida's power system with high PV penetration and battery storage technologies. They considered total residential, commercial, and utility loads, along with flexible options like battery storage and demand response to increase PV penetration and reduce carbon emissions. Aniello et al. performed a techno-economic assessment for German households in \cite{aniello2021micro} by considering PV and battery storage systems. They used various financial metrics such as NPV, cash flows, simple payback, and internal rate of return. Alipour et al. \cite{alipour2022exploring} investigated the motivations and concerns of residential customers in Queensland, Australia, regarding PV and battery adoption. They found that upfront costs and reliability concerns were key factors. Tervo et al. in \cite{tervo2018economic} conducted an economic analysis of residential PV systems coupled with Li-ion battery storage across 50 U.S. states. They discovered that optimally sized PV-battery systems with net-metering were more cost-effective than stand-alone PV systems. Mohamed et al. presented a robust techno-economic tool for UK residential PV-battery systems in \cite{mohamed2021comprehensive}. Their analysis considered real-time battery control, capacity degradation, and investment profitability using metrics like NPV and return on investment. Koskela et al. evaluated the profitability of optimally sized PV and battery storage systems for apartment buildings and detached houses in Finland in \cite{koskela2019using}. They also considered economic analysis regarding equipment sizing in response to tariff and incentive changes in power distribution systems.  William et al.  \cite{william2022enviro} analyzed the potential of energy-efficient building solutions in various hot and humid climates through parametric analysis. Using an environmental assessment, energy consumption was reduced by installed PV, and indoor thermal discomfort was mitigated. Way et al. \cite{way2022empirically} generated empirically validated cost forecasts for various technologies impacting the energy transition period. They explored different energy transition scenarios and found that previous cost projections overestimated the costs of renewable technologies. In another study, Kunwar et al. \cite{kunwar2023performance} investigated the function and efficacy of active insulation systems in the context of enhancing the efficiency of residential buildings, with the primary objective of mitigating energy loss and reducing peak demand.
Further information on Net Zero Emission Buildings and strategies for retrofitting existing buildings, increasing renewable penetration at the customer level, and relevant policies can be found in \cite{ohene2022review}, \cite{zhang2022grid}, and \cite{ahmed2022assessment}.\par

\vspace{2mm}

Previous research studies have primarily focused on the economic analysis of residential households equipped with PV and battery storage systems, utilizing financial indexes. Some of these studies have addressed aspects like PV and battery sizing, energy efficiency improvements, and more. However, a comprehensive analysis of the economic and technical benefits of NZEBs over the next 30 years, encompassing PV, energy efficiency enhancements, and stationary battery storage systems, is lacking. A comprehensive analysis should consider various aspects, such as monthly savings for residents, the timeline at which residential renewable electricity could become more cost-effective than purchasing electricity from the grid, and the potential for peer-to-peer energy exchanges among residents, all while ensuring reliable energy supply during emergencies. Such an analysis should explore the advantages NZEBs offer to utilities, including: reducing overall grid load and increasing renewable capacity for hydrogen (H2) production which can play a crucial role in decarbonizing both the power and transportation sectors. There is also a gap in research that focuses on using real-world energy consumption from existing and retrofitted houses in the state of Florida.

In this paper, the following questions will be addressed: \par
\begin{itemize}

\item What is the monthly savings for adding solar to the roof of the average home in Florida to make it a Net Zero Electricity Residence (function of the year of solar installation)?
\item What is the monthly savings for adding solar and energy efficiency improvements to the average new home construction in Florida to make it a Net Zero Electricity Residence (function of the year of new home construction)?
\item What year will Florida net zero homeowners be able to cost-effectively store all the electricity they generate in their own batteries which can provide backup power when the electric grid is not available? What year will this homeowner’s cost be less than the Utility cost of shipping solar to the home?
\item What year will Florida residents pay less per month on their energy needs for their residences and their fuel for transportation by adding a single EV, while still having reliable backup power from their electric vehicle through the vehicle to home (V2H) when the electric grid is not available?  
\item When could Florida Utilities make more profit by making hydrogen than delivering solar electricity to customers?
\end{itemize}
\vspace{2mm}

To address these questions, this paper examines the monthly cost savings of adding a 9.5 kW solar system to an existing Florida residence, transforming it into an average Net-Zero Energy Residence. To achieve an average level of grid independence, the addition of 42 kWh of battery storage is required. The paper compares the monthly cost savings of owning PV on the roof with zero energy storage, 50\%, and 100\% effective storage, both with and without a Federal income tax credit (ITC) of 30\%, against the cost of purchasing electricity from the grid for the years 2020 to 2050. The monthly cost savings of a new Orlando net-zero home constructed with efficiency improvements, PV installation, and battery storage, are compared to the cost of buying electricity from the grid. The paper also explores the monthly cost savings of a Net-Zero residence using an average 68.7 kWh electric vehicle as emergency backup power through Vehicle-to-Home (V2H) technology. Lastly, the technical and economic benefits of NZEBs in increasing residential renewable penetration, effectively eliminating residential load on the grid are discussed. These benefits create opportunities for both residents and utilities, enabling Florida to achieve net-zero emissions at a minimal cost. Residents can benefit from monthly savings, engage in peer-to-peer energy exchanges, and have access to reliable backup power. Utilities stand to gain by reducing transmission losses, operating residential PV and batteries as virtual power plants, and using utility-scale solar more efficiently to facilitate fast charging of vehicles and hydrogen (H2) production, thus decarbonizing the power and transportation sectors.\par
\vspace{2mm}

The remaining sections of the paper are presented and discussed as follows: First, the forecasts for the installed cost of residential PV systems, along with levelized cost of electricity (LCOE) forecasts which include the electricity cost breakdown based on generation, transmission, distribution, and profit \& taxes for PV-only scenarios will be presented. Following that, Florida's electricity consumption for both existing and new Orlando code-compliant residences, as well as their respective monthly savings will be discussed. Next, the installed cost of residential PV + battery systems, along with the associated monthly savings for Florida residences will be presented. Simulation results for both existing and new code-compliant homes, with and without efficiency improvements, scenarios with and without the ITC, and case studies with and without V2H capabilities will be presented. In the final section, the benefits of NZEBs on the total electricity demand of utilities as well as assessing the impacts of the future cost of hydrogen (H2) production for both the power and transportation sectors will be discussed. Finally, the paper will conclude with a summary of findings.

\section{Installed Cost of Residential PV from 2020 to 2050} \label{InstalledCost_PV}

\begin{figure}
\centering
\footnotesize
	\includegraphics[width=5.5in]{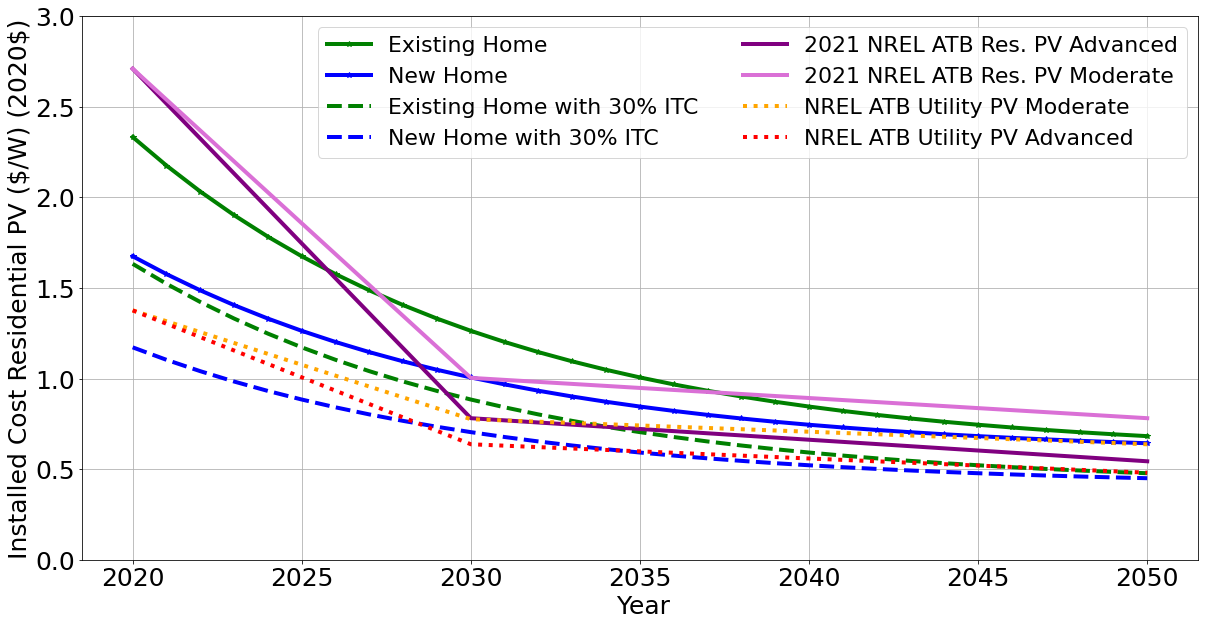}
	\caption{Installed U.S. Residential PV Price Curve in \$ per Wpdc }
    \label{PVInstalled_residential}
\end{figure}

Figure \ref{PVInstalled_residential} shows the cost of residential rooftop solar installations in the U.S. for existing homes, measured in \$ per Wpdc (2020\$), using data from the National Renewable Energy Laboratory (NREL) Annual Technology Baselines (ATB 2021) \cite{NRELATB}. In the figure, the purple and orchid lines represent the NREL ATB data. The solid green curve, which has been superimposed on the graph, was developed by the authors. It assumes a price leveling to \$0.7 in 2050, a value situated between the two ATB lines projected for 2050. The solid green curve is conservative when compared to the ATB curves between 2027 and 2037. The installed cost in 2020 of \$2.26/W (2020\$) reflects the cost gathered from Solar United Neighbors of Florida's 1600 cooperative customers. As a result, the green curve from 2020 to 2026 falls below the two ATB lines. This green curve will be utilized throughout this paper as the reference for the installed system price for residential rooftops in existing homes. For new home construction costs of installed PV, represented by the solid blue curve, have reduced labor and other soft installation costs at the time of construction. This reduction assumes that the installation costs for new construction will be equivalent to those for existing construction five years into the future. By 2050, the installed costs for both existing and new home construction converge, as seen by the merging of the green and blue curves.


\begin{figure}
\centering
\footnotesize
	\includegraphics[width=5.5in]{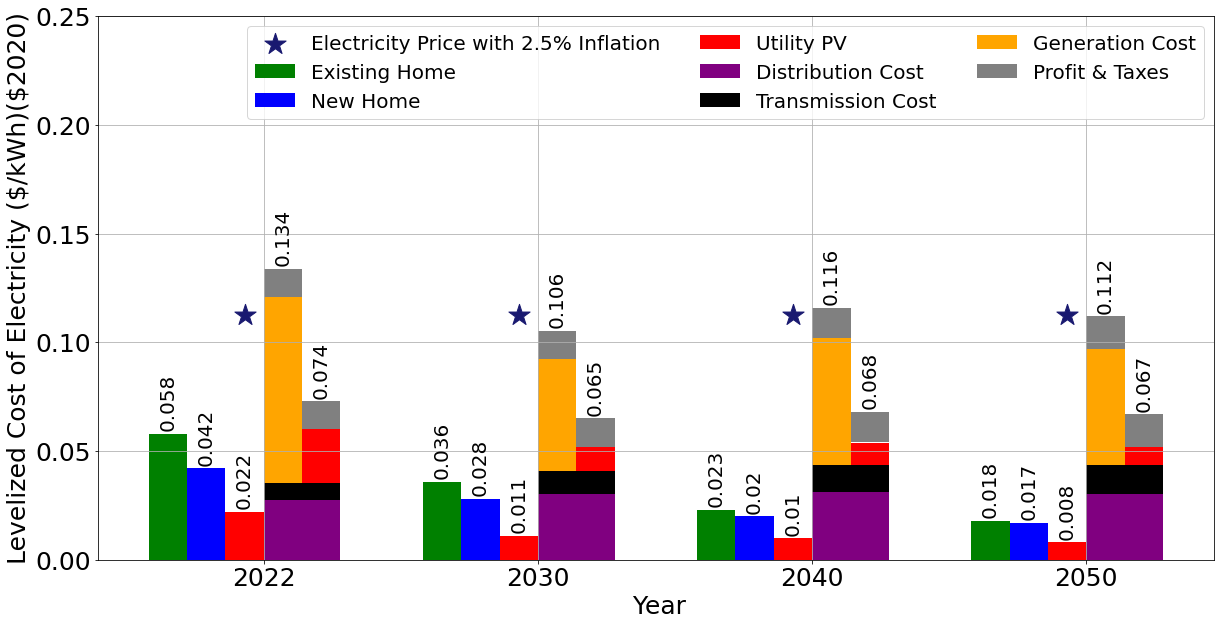}
	\caption{Florida Levelized Cost of Electricity; Two Residential PV Homes, Utility PV, and Residential Electricity “out of the wall”, with No Income Tax Credit}
    \label{LCOE_PV_Residential}
\end{figure}

\section{LCOE of Residential PV from 2020 to 2050}

Figure \ref{LCOE_PV_Residential} presents the unsubsidized (no ITC) LCOE for residential rooftop solar, utility solar, distribution, transmission, generation, and profit \& taxes in Florida (The same figure that includes ITC is shown in Figure \ref{LCOE_PV_Residential1} in the Appendix section). This analysis uses the installed cost of solar in peak watts and assumes a PV effectiveness of 1,400 kWh/kWpdc per year. In 2020 the cost of residential electricity out of the wall was 11.3 ¢ per kWh \cite{EIAprice} which is the lowest cost for residential customers in 2020\$ for 1990 to 2023 (as shown in Figure \ref{Historic_prices} of the Appendix). The “stars” represent 11.3 ¢ per kWh in \$2020 for each year assuming that the price of electricity increases at the 2.5\% general inflation rate. For all calculations in future years the 2020 cost of electricity out of the wall, 11.3 ¢ per kWh in 2020\$ does not change with time as the price is assumed to increase at the general rate of inflation, 2.5\%, which then yields very conservative values compared to the cost of electricity out of the wall in 2023 which is 15.36 ¢ per kWh in 2023\$ or 14.3 ¢ per kWh in 2020\$. In 2022, the unsubsidized LCOE for existing residential rooftop PV is 5.8 ¢ per kilowatt-hour. By 2030, it drops to 3.6 ¢ per kilowatt-hour, and further decreases to 2.3 ¢ per kilowatt-hour in 2040 and 1.8 ¢ per kilowatt-hour in 2050, all values expressed in 2020\$. Similarly, for new homes constructed with PV installed, the LCOE in 2022 is 4.2 ¢ per kilowatt-hour, and reduces to 2.8 ¢ per kilowatt-hour by 2030, 2 ¢ per kilowatt-hour in 2040, and 1.7 ¢ per kilowatt-hour in 2050, all in 2020\$. The Utility PV LCOE, represented by the red column, is directly sourced from NREL ATB 2021 \cite{NRELATB}. In 2022, the Utility PV LCOE is 3.1 ¢ per kilowatt-hour, decreasing to 1.6 ¢ per kilowatt-hour by 2030, 1.4 ¢ per kilowatt-hour in 2040, and 1.2 ¢ per kilowatt-hour in 2050, all in 2020\$. For each year shown, the LCOE of existing homes is higher than that of new homes constructed to the IECC standard, which, in turn, is higher than the utility LCOE. However, these differences diminish over time. Reference \cite{EIAprice} forecasts the residential electricity price from Florida utilities from 2022 to 2050, encompassing distribution (purple), transmission (black), generation (orange), and profit \& taxes (grey) costs. The last column for each year replaces the generation LCOE in the fourth column with the Utility PV generation LCOE.  While generation costs decrease annually, distribution and transmission costs rise. In 2022, the residential PV LCOE for existing homes and new homes is 50\% and 36\% of the total electricity "cost out of the wall" respectively. By 2030, these percentages decrease to 34\% and 26\%, respectively. The final column for each year presents the utility PV LCOE alongside distribution, transmission, generation, profit \& tax (grey) costs. This column reflects the cost to the utility for producing and delivering solar electrons to residential customers. In 2022, while utilities produce solar PV electricity at a lower cost than residents, utility solar electricity delivered to homes is more expensive. By 2030, residents can generate solar electricity at a lower cost than the combined distribution and transmission costs. As a result, by 2030, utilities may allocate a significant portion of their solar electricity to facilitate fast charging of vehicles and for hydrogen production via electrolysis, which could have greater value than simply delivering solar electrons over long distances. This topic will be explored further in this paper. 

\section{Gasoline Equivalent of Residential PV LCOE from 2020 to 2050}

\begin{figure}
\centering
\footnotesize
	\includegraphics[width=5in]{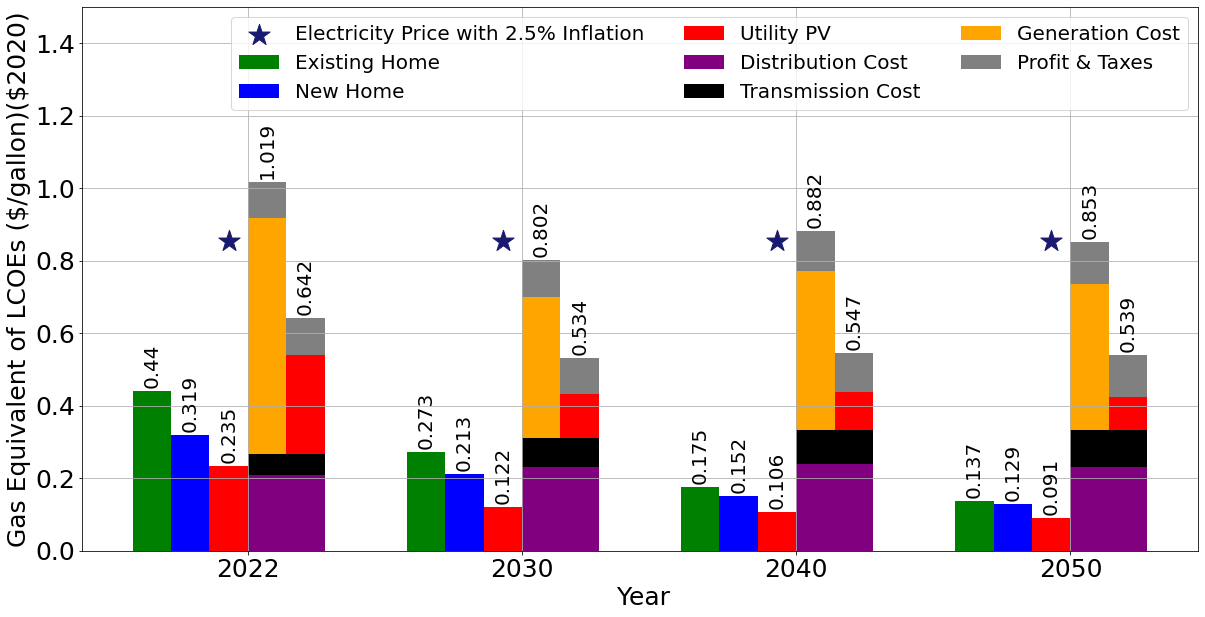}
	\caption{Gas Equivalent of LCOEs (\$/gallon) for PV Only Scenario without 30\% PV ITC.}
    \label{gas_equi_PVonly}
\end{figure}

Figure \ref{gas_equi_PVonly} is similar to Figure \ref{LCOE_PV_Residential} with the LCOE but here the LCOE is converted to gasoline equivalent in dollars per gallon.  A similar figure to Figure \ref{gas_equi_PVonly} that includes the ITC is shown in Figure \ref{gas_equi_PVonly_ITC} in the Appendix section. The average EV has a range of 220 miles, has a 68.7 kWh battery which means the efficiency is 3.20 miles/kWh.  The average gasoline vehicle gets 24.2 mpg \cite{USDOEEERE}.  So the gasoline equivalent in \$ per gallon can now be calculated using the price of electricity in \$/kWh multiplied by 24.2 mpg divided by 3.20 miles/kWh. The LCOE for rooftop solar in the existing home in 2022 (green column of Figure \ref{LCOE_PV_Residential}) is 5.8 ¢ per kWh which is equivalent to \$0.44/gallon (= \$0.058 * 24.2/3.2).  Since the average price for gasoline in 2023 is \$3.56 per gallon in 2023\$ \cite{GasFLPrice}  or \$3.16 per gallon in 2020\$, the gasoline equivalent of PV on the roof is one-seventh the cost of gasoline.  It is clear that for all cases of PV electricity generation the gasoline equivalent cost in Figure \ref{gas_equi_PVonly} is always less than \$0.65 per gallon.

\section{Florida Electricity Consumption for Existing Residences}\label{FLA_ElecConcumption_Existig}

\begin{figure}
\centering
\footnotesize
	\includegraphics[width=6.5in]{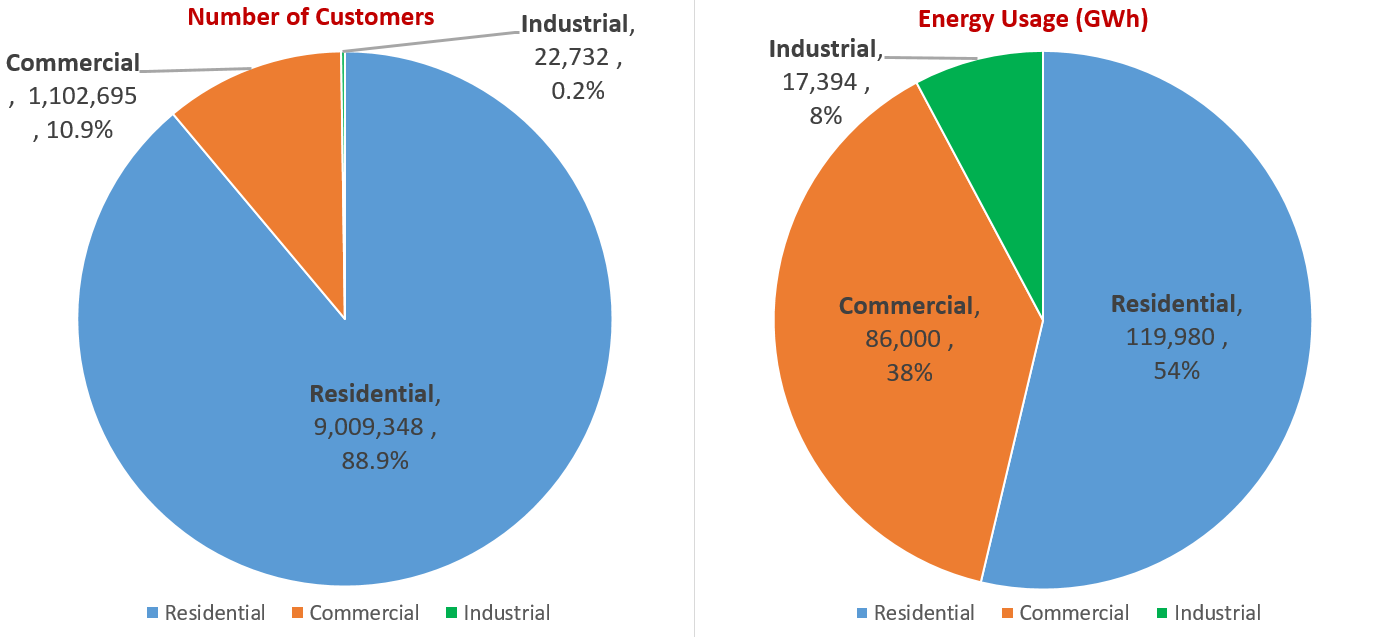}
	\caption{Florida Electric Customer Composition 2020}
    \label{FL_Electricity_Piecharts}
\end{figure}

Figure \ref{FL_Electricity_Piecharts} illustrates the composition of Florida's electric customer base in 2020, sourced from reference \cite{FLAPSC}. The left pie chart reveals that residential customers account for 89\% of the total customer base. Meanwhile, the right pie chart demonstrates that 54\% of the electricity is consumed in residences, with 39\% used by commercial establishments. This indicates that more than 90\% of Florida's electricity consumption occurs in buildings. To calculate the average electricity consumption per customer, the energy usage from the right pie chart is divided by the number of customers in the left pie chart. For residential customers, this averages out to 13,300 kWh per year (approximately 1,100 kWh per month or 36 kWh per day). With an electricity cost of 11.3 ¢ per kilowatt-hour, this results in a monthly bill of \$124. Commercial customers, on the other hand, consume an average of 77,000 kWh per year, while industrial customers use an average of 760,000 kWh per year. If each residential customer were to have access to 9.5 kW of PV (assuming 1,400 kWh of solar electricity generated per kW of PV annually), each residential customer would have a "net-zero" energy home.
 
\section{Monthly Saving for Net Zero Existing Florida Residences}\label{MonthlySaving_PV}

Using 9.5 kW of PV to produce the 13,300 kWh/year in electricity consumption and the values from Figure \ref{PVInstalled_residential} shows the unsubsidized monthly savings (green bar) and the subsidized monthly savings (Federal ITC at 30\%; diagonal patterned green bar) to make an average home in Florida a Net Zero Electricity Residence. The customer saves money every month without the ITC and saves even more per month with the ITC.  A Net Zero Electric Residence in Florida saves the customer money  TODAY.  In 2020 the unsubsidized Net Zero Existing Residence has a monthly savings of \$18 over paying the electric bill for the home without solar.  If the solar was installed in 2030 the monthly savings would be \$52 and in 2050 the savings would be \$71 per month all in 2020\$.  If the 30\% ITC is available the monthly savings are \$40 in 2020, \$64 in 2040, and \$77 in 2050, per month in 2020\$.

\begin{figure}
\centering
\footnotesize
	\includegraphics[width=5.5in]{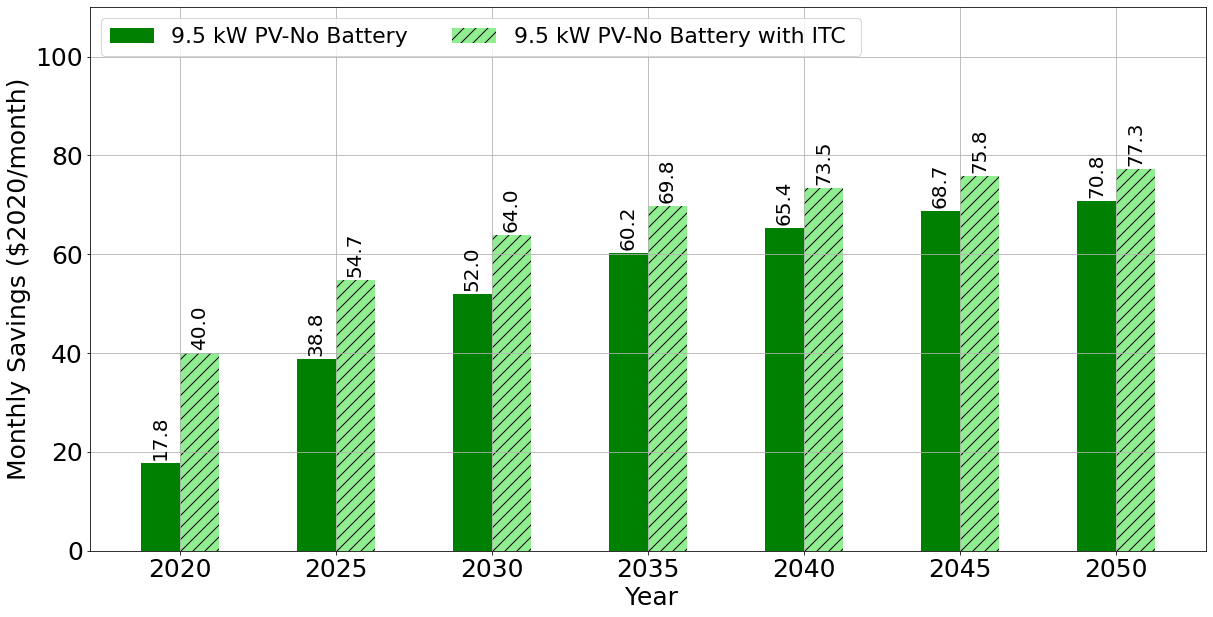}
	\caption{Monthly Savings for Florida Existing Residence from 2020 to 2050}
    \label{Monthly_saving_PV}
\end{figure}

\section{Florida Electricity Consumption New Home Construction}

In Florida, today's new home construction must have better energy efficiency than the minimum requirements of the International Energy Conservation Code (IECC) home. To assess the cost-effectiveness of high-performance homes that significantly surpass these minimum requirements, information from the Energy-Gauge® USA (v.4.0.00) \cite{EnergyGauge} was used. The study's objective was to determine the highest level of energy efficiency that remains cost-effective for consumers. The analysis focused on the cost-effectiveness of the entire package of energy-efficient measures rather than individual measures. A one-story 2,000 ft², 3-bedroom homes, and two-story 2,400 ft², 3-bedroom homes in thirteen representative cities \cite{center2015maximum} with typical meteorological year (TMY) data, representing eight IECC climate zones were considered. The energy performance of these high-performance Improved Homes was compared to both the IECC Code Homes and the SSPC 90.2 Reference Homes. Simulations are conducted for each home, considering both best-case and worst-case home orientations. Improvements were made to the IECC Orlando homes to ensure that the cost-effectiveness of the improved homes resulted in a savings-to-investment ratio (SIR) greater than 1.00. This determination allowed the maximum efficiency level to be identified that remains cost-effective for consumers.

\begin{table}[]
\centering
\caption{Orlando IECC Homes (code homes) with and without Efficiency Improvements}
\label{CodeHomeTable}
\begin{tabular}{|l|cccc|cccc|}
\hline
\multicolumn{1}{|c|}{} &
  \multicolumn{4}{c|}{\textbf{IECC Homes}} &
  \multicolumn{4}{c|}{\textbf{ERI Homes}} \\ \cline{2-9} 
\multicolumn{1}{|c|}{\multirow{-2}{*}{\textbf{Case}}} &
  \multicolumn{1}{c|}{\textbf{kWh/y}} &
  \multicolumn{1}{c|}{\textbf{Th/y}} &
  \multicolumn{1}{c|}{\textbf{\$/yr}} &
  \textbf{HERS} &
  \multicolumn{1}{c|}{\textbf{kWh/y}} &
  \multicolumn{1}{c|}{\textbf{Th/y}} &
  \multicolumn{1}{c|}{\textbf{\$/yr}} &
  \textbf{HERS} \\ \hline \hline
\textbf{1-sty Best Case} &
  \multicolumn{1}{c|}{11417} &
  \multicolumn{1}{c|}{0} &
  \multicolumn{1}{c|}{\$1427} &
  79 &
  \multicolumn{1}{c|}{7773} &
  \multicolumn{1}{c|}{0} &
  \multicolumn{1}{c|}{\$972} &
  50 \\ \hline
\textbf{1-sty Worst Case} &
  \multicolumn{1}{c|}{11522} &
  \multicolumn{1}{c|}{0} &
  \multicolumn{1}{c|}{\$1441} &
  81 &
  \multicolumn{1}{c|}{7839} &
  \multicolumn{1}{c|}{0} &
  \multicolumn{1}{c|}{\$980} &
  50 \\ \hline
\textbf{2-sty Best Case} &
  \multicolumn{1}{c|}{12644} &
  \multicolumn{1}{c|}{0} &
  \multicolumn{1}{c|}{\$1581} &
  78 &
  \multicolumn{1}{c|}{8653} &
  \multicolumn{1}{c|}{0} &
  \multicolumn{1}{c|}{\$1082} &
  49 \\ \hline
\textbf{2-sty Worst Case} &
  \multicolumn{1}{c|}{12759} &
  \multicolumn{1}{c|}{0} &
  \multicolumn{1}{c|}{\$1595} &
  79 &
  \multicolumn{1}{c|}{8743} &
  \multicolumn{1}{c|}{0} &
  \multicolumn{1}{c|}{\$1093} &
  49 \\ \hline
{\color[HTML]{CB0000} \textbf{Averages}} &
  \multicolumn{1}{c|}{{\color[HTML]{CB0000} \textbf{12086}}} &
  \multicolumn{1}{c|}{{\color[HTML]{CB0000} \textbf{0}}} &
  \multicolumn{1}{c|}{{\color[HTML]{CB0000} \textbf{\$1511}}} &
  {\color[HTML]{CB0000} \textbf{79}} &
  \multicolumn{1}{c|}{{\color[HTML]{CB0000} \textbf{8252}}} &
  \multicolumn{1}{c|}{{\color[HTML]{CB0000} \textbf{0}}} &
  \multicolumn{1}{c|}{{\color[HTML]{CB0000} \textbf{\$1032}}} &
  {\color[HTML]{CB0000} \textbf{50}} \\ \hline
\end{tabular}
\end{table}

Table \ref{CodeHomeTable} reveals that the average IECC home in Orlando consumed 12,086 kWh per year with a HERS rating of 79. In contrast, the more energy-efficient ERI home used 8,252 kWh per year with a HERS rating of 50. These efficiency improvements resulted in a 31.7\% electricity savings at an initial cost of \$5,889, equating to \$1.358 per kWh saved and a savings-to-investment ratio of 1.14. 

\begin{figure}
\centering
\footnotesize
	\includegraphics[width=5.5in]{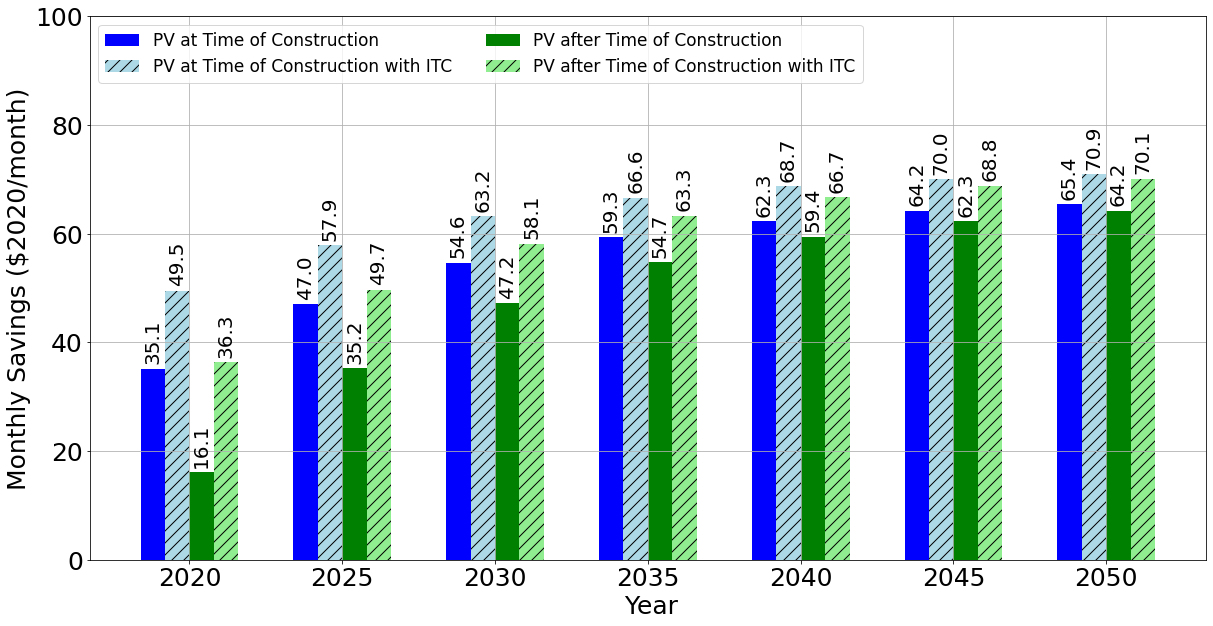}
	\caption{Monthly Savings for Florida IECC Home Residence 2020 to 2050}
    \label{Monthly_saving_CodeHomes}
\end{figure}

Figure \ref{Monthly_saving_CodeHomes} shows the monthly savings for a Net Zero Code Home, which has a HERS rating of 0 and employs an 8.63 kW PV system to generate 12,806 kWh/year of electricity. These monthly savings were calculated considering PV installation either during home construction or after construction, with and without the Investment Tax Credit (ITC). In all cases of Net Zero Code Orlando Residences, homeowners experience monthly savings. In 2020, an unsubsidized Net Zero Code Residence with PV installed during construction (represented by the blue column) offers a monthly savings of \$35 compared to paying the electric bill without solar. If a new home is constructed with solar in 2030, the monthly savings increase to \$55, and in 2050, they reach \$65, all in 2020\$. With the 30\% ITC available (indicated by the diagonal blue column), monthly savings are \$50 for solar construction in 2020, \$63 in 2030, and \$71 in 2050, per month in 2020\$. For solar installations after construction in the years 2020, 2030, and 2050, the monthly savings would be \$16, \$47, and \$64 in 2020\$, respectively. If the 30\% ITC is available for post-construction solar installation (represented by the upward diagonal patterned green column) in these years, monthly savings increase to \$36, \$58, and \$70 in 2020\$, respectively.

\begin{figure}
\centering
\footnotesize
	\includegraphics[width=5.5in]{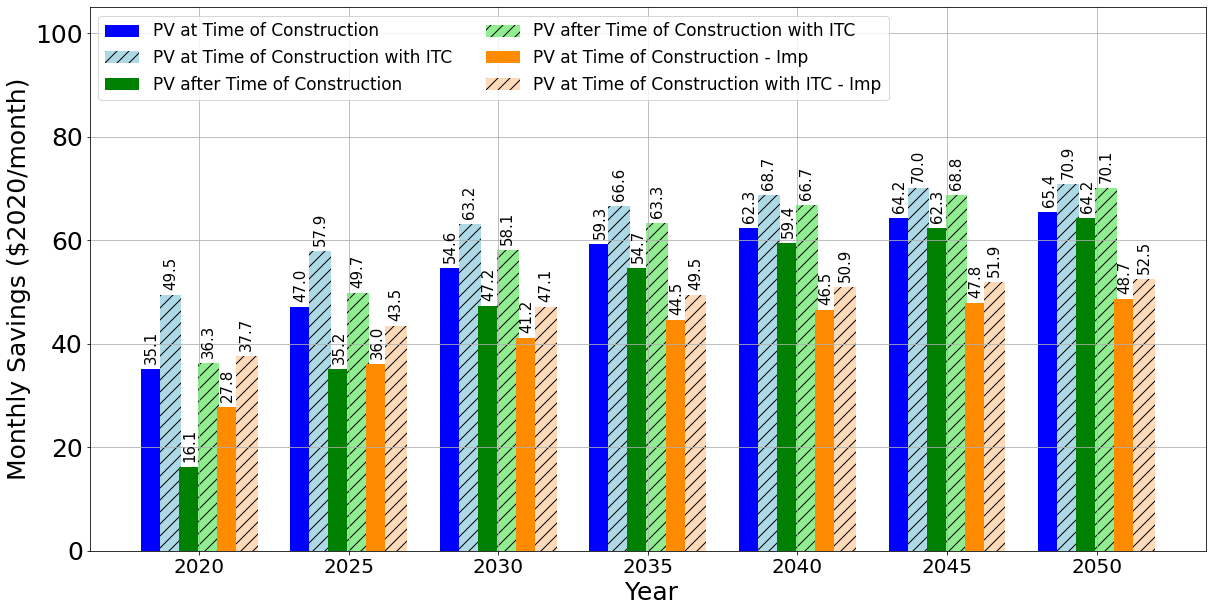}
	\caption{Monthly Savings for IECC Home Residence w/ and w/o Efficiency Improvements from 2020 to 2050}
    \label{Monthly_saving_Imp}
\end{figure}

Figure \ref{Monthly_saving_Imp} presents the Monthly Savings for two types of homes: one is a Code home equipped with an 8.63 kW PV system (the same blue and green columns as shown in Figure \ref{Monthly_saving_CodeHomes}), and the other is a Code home with 31.7\% more efficient improvements and a 5.89 kW PV system to achieve Net Zero status (represented by the orange columns). Both homes have HERS ratings of 0, and all Net Zero Electric Residences in Figure \ref{Monthly_saving_Imp} result in monthly savings for homeowners. In 2020, the unsubsidized Net Zero Code Residence, with both efficiency improvements and solar installed during construction (orange column), yields a monthly savings of \$28. If the 31.7\% more efficiency improvements and 5.89 kW of PV are installed during construction in 2030 and 2050, the monthly savings would be \$41 and \$49 in 2020\$, respectively. When the 30\% Investment Tax Credit (ITC) is available for the more efficient than code residence with solar installed at the time of construction (indicated by the diagonal patterned orange column) in 2020, 2030, and 2050, the monthly savings would be \$38, \$47, and \$53 in 2020\$, respectively. The highest monthly savings occur with the code home where the PV is installed at the time of construction (blue column). Efficiency improvements become cost-competitive with PV when the price of PV exceeds \$2.25/W in 2020 or \$2.12/W in 2025. Otherwise, PV is more cost-effective than improved efficiency, as long as net metering is in effect.

Net metering (NEM) has made rooftop solar systems a valuable investment and has facilitated the development of net-zero energy homes and buildings. However, it poses challenges to traditional electric utilities, as reduced demand from self-generating consumers can lead to increased rates for others, covering the fixed grid investment costs. Some states are exploring demand charges, time-of-use pricing structures, and monthly surcharges. It's important for all parties to recognize that grid costs should be uniform for all customers, whether they have solar or not. In states with deregulated grids, customers pay one utility for grid use and another for energy costs. Regardless of deregulation, transparent payments for grid charges and electric energy are essential for Florida customers to make wise investments.

\begin{table}[]
\centering
\caption{2020 Financial Indexes for Florida Residents with PV Only}
\label{PVonly_Financial_Table11}
\begin{tabular}{|c|c|c|c|c|c|c|} 
\hline
\textbf{\begin{tabular}[c]{@{}c@{}}Case\\ (PV Only)\end{tabular}} &
  \textbf{\begin{tabular}[c]{@{}c@{}}Existing \\ Home\end{tabular}} &
  \textbf{\begin{tabular}[c]{@{}c@{}}New \\ Home\end{tabular}} &
  \textbf{\begin{tabular}[c]{@{}c@{}}New \\ Home\\ with Imp.\end{tabular}} &
  \textbf{\begin{tabular}[c]{@{}c@{}}Existing \\ Home \\ with ITC.\end{tabular}} &
  \textbf{\begin{tabular}[c]{@{}c@{}}New Home \\ with ITC.\end{tabular}} &
  \textbf{\begin{tabular}[c]{@{}c@{}}New Home \\ with ITC \& Imp.\end{tabular}} \\ \hline \hline
\rowcolor[HTML]{EFEFEF} 
\textbf{\begin{tabular}[c]{@{}c@{}}NPV\\ (\$)\end{tabular}}   & 5319  & 11581 & 9079  & 11977 & 16361 & 12343\\ \hline
\textbf{\begin{tabular}[c]{@{}c@{}}IRR\\ (\%)\end{tabular}}   & 6.51  & 10.12 & 8.67 & 10.42 & 14.95 &  11.14 \\ \hline
\rowcolor[HTML]{EFEFEF} 
\textbf{\begin{tabular}[c]{@{}c@{}}SIR\\ \;\end{tabular}}      & 1.38  & 1.88  &  1.7& 1.93  & 2.65 & 2.07  \\ \hline
\textbf{\begin{tabular}[c]{@{}c@{}}SPB\\ (Year)\end{tabular}} & 15.69 & 11.27 &  12.28 & 10.98   & 7.89 & 9.97 \\ \hline
\end{tabular}

\end{table}



Table \ref{PVonly_Financial_Table11} shows different financial indexes including net present value (NPV), internal rate of return (IRR), saving-to-investment ratio (SIR), and simple payback (SPB) for Florida residents with PV on the roof. Six different scenarios are considered; Florida residents who equip their houses with PV after the construction (existing home), at the time of construction (new home), and new homes with 31.7\% efficiency improvement without 30\% PV ITC. The highest IRR and SPB belong to new homes with 14.95\% and 7.89 years, respectively if the owner benefits from the PV ITC.

While cheap residential solar presents challenges for utilities, the advent of affordable energy storage will prove to be even more disruptive. The combination of solar energy, wall-mounted batteries, and electric vehicle-to-home technology empowers customers to arbitrage variable rate designs or demand charges created by utilities, rendering such rate structure changes ineffective in mitigating utility load losses. Customer storage enables individuals to utilize a significant portion of their solar generation to meet their own electricity needs, allowing them to retain nearly the full retail value of their solar production. This trend raises concerns about widespread partial grid defection, where customers remain connected to the grid for 24/7 reliability but generate 80 to 90 percent of their own energy and use storage to optimize their solar for personal consumption. In Florida, consumers are poised to take control of their solar-generated electricity by storing it in residential batteries and electric vehicles, as it will soon become more cost-effective than purchasing electricity from the grid. The pivotal question is: when will Florida's consumers make this transition? To answer this, the forecasts for the installed battery costs must be known.

\section{Battery Costs and Projections}\label{InstalledCost_PVBattery}
\begin{figure}
\centering
\footnotesize
	\includegraphics[width=5.7in]{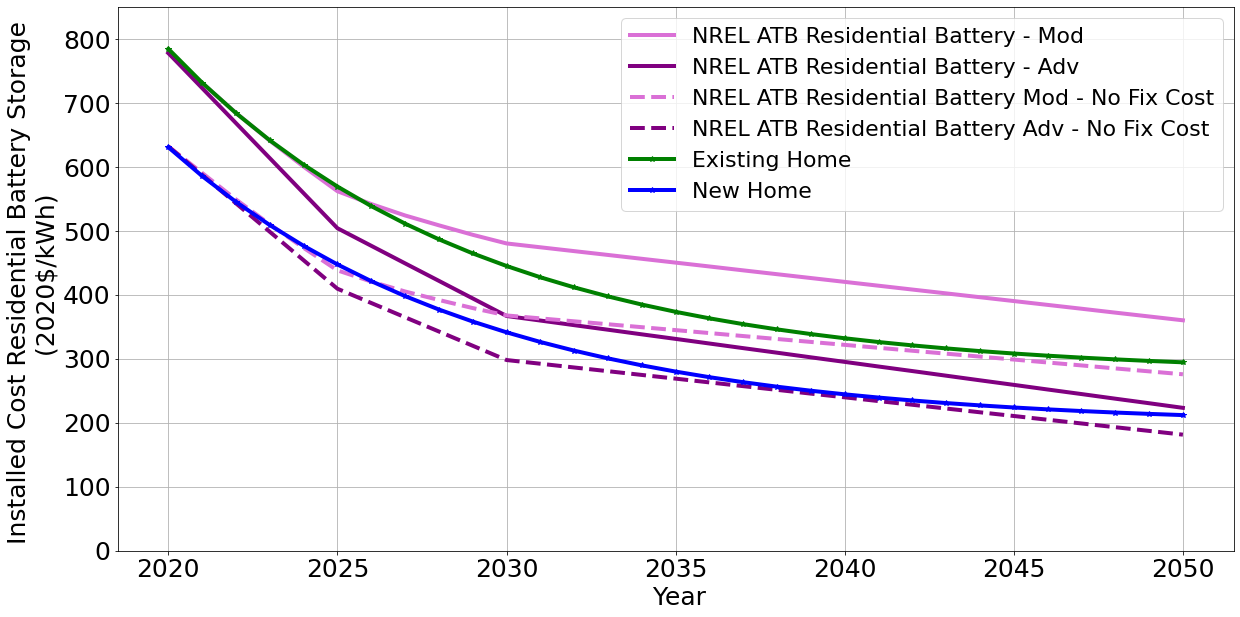}
	\caption{Installed Residential Battery Price Curve}
    \label{Installed_BESCost}
\end{figure}

The storage of solar electricity has become essential as it's true that the sun doesn't shine all the time. Much like solar PV generation, battery technology has seen rapid cost reductions over the past ten years, evolving from being considered "too expensive" to becoming the "most cost-effective" option for utility-scale solar with four hours of battery storage. From a utility business perspective, large-scale solar and energy storage solutions are poised to replace natural gas combined-cycle power plants and peaking natural gas power plants. The development of multiple Gigafactories (facilities producing over 100 GWh/year of battery packs) worldwide is expected to drive further cost reductions, outpacing previous projections. In \cite{TeslaSunrun}, the cost of several installed home battery systems, including Tesla and Sunrun, is provided. The Tesla Powerwall 2, for example, is a 13.5 kWh battery with a 10-year warranty, suitable for standalone energy storage or integration with PV panels. Figure \ref{Installed_BESCost} displays a price forecast in 2020\$ per kWh for residential energy storage, sourced from the NREL ATB \cite{NRELATB}. The solid and dashed purple and orchid curves represent the cost of residential storage, both with and without fixed costs, for moderate and advanced forecasts. The solid green curve represents the price forecast used by the authors for residential batteries installed in existing houses. This curve aligns well with the NREL ATB moderate curve from 2020 to 2025 and then converges with the average of the moderate and advanced curves. The solid blue line represents the price forecast used by the authors for residential batteries installed during the construction of houses. This curve tracks the NREL curves, assuming no fixed installation costs as the batteries are integrated at the time of construction. In 2020, the cost of residential energy storage installed in existing construction (green curve) is \$784 per kWh. By 2025, it is expected to decrease to \$569 per kWh installed, further dropping to \$445 per kWh in 2030 and \$332 per kWh in 2040. For energy storage installed during residential construction (blue curve), the cost in 2020 is \$631 per kWh, projected to decrease to \$447 per kWh in 2025, \$341 per kWh in 2030, and \$244 per kWh in 2040.\par

\begin{figure}
\centering
\footnotesize
	\includegraphics[width=5.5in]{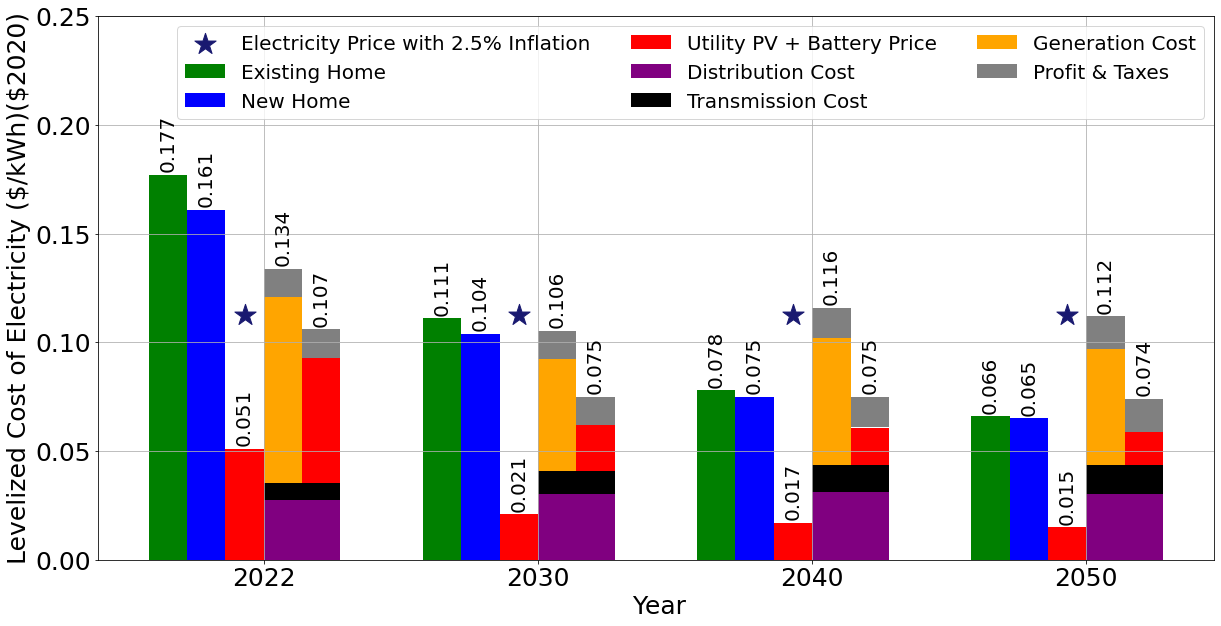}
	\caption{Florida Residential PV + Battery (100\% effective storage) and Utility PV + Battery (4-hour) "out of the wall" Cost of Electricity, without the ITC for PV and Battery}
    \label{LCOE_PV_Batt_Comparison}
\end{figure}


Now the following question can be addressed: In what year will Florida's consumers take control of their own solar-generated electricity by storing it in their residential batteries, as it will become more cost-effective than purchasing electricity from the grid?



\section{LCOE of Residential PV + Battery System from 2020 to 2050}

Figure \ref{FL_Electricity_Piecharts} showed that the average Florida residence consumes 13,300 kWh/year in electricity. To achieve Net Zero Electricity status, an average residence requires a 9.5 kW solar PV system. However, to be completely grid-independent on average, considering occasional cloudy days in Florida, one needs more than 36.5 kWh (13,300 kWh/year divided by 365 days/year) for daily load coverage. Taking into account the gradual decrease in the performance of both the PV system and the battery over the 25-year lifespan, a total of 42.2 kWh of storage is required to ensure grid independence for the average residence over 25 years.

Figure \ref{LCOE_PV_Batt_Comparison} displays the Levelized Cost of Electricity (LCOE) without ITC for PV + Battery systems for different years, including the existing home (green column, featuring an average grid-free 9.5 kW PV + 42.2 kWh battery), the new IECC home + PV + battery installed at the time of construction (blue column, with an average grid-free 5.9 kW PV + 26.2 kWh battery), and a Utility PV system with a 4-hour peak shifting battery (red column, consisting of 100 MWAC PV + 40 MW/160 MWh-hour lithium-ion battery). Figure \ref{LCOE_PV_Batt_Comparison1} in the Appendix section shows this same figure with the ITC

In 2022, the unsubsidized LCOE for existing residential rooftop PV + 100\% battery is 17.7 ¢ per kilowatt-hour. By 2030, the LCOE will decrease to 11.1 ¢ per kilowatt-hour, and in 2040, it's 7.8 ¢ per kilowatt-hour. In 2050, the LCOE will reach 6.6 ¢ per kilowatt-hour, all values in 2020\$. For the new IECC home with efficiency improvements + PV + 100\% battery installed at the time of construction, the PV LCOE is 16.1 ¢ per kilowatt-hour in 2022. By 2030, it drops to 10.4¢ per kilowatt-hour, and in 2040, it's 7.5¢ per kilowatt-hour. In 2050, the LCOE is 6.5¢ per kilowatt-hour, all values in 2020\$. The Utility PV + 4-hour battery LCOE (red column) was obtained from Reference NREL ATB 2021 \cite{NRELATB}. In 2022, the Utility PV + battery LCOE is 5.1 ¢ per kilowatt-hour. By 2030, it decreases to 2.1 ¢ per kilowatt-hour, and in 2040, it's 1.7 ¢ per kilowatt-hour. In 2050, the LCOE will reach 1.5 ¢ per kilowatt-hour, all values in 2020\$. As previously mentioned in Figure \ref{LCOE_PV_Residential}, reference \cite{EIAprice} forecasts residential electricity prices from Florida utilities from 2022 to 2050, considering distribution (purple), transmission (black), generation (orange), and profit \& taxes (grey) costs. While generation costs decline year by year, distribution and transmission costs increase over time. In 2030, the LCOE for the two homes that are on average grid-free becomes comparable to the electricity cost out of the wall (10.6 ¢ per kilowatt-hour). By 2040, the LCOE for these homes is approximately 4 ¢ per kilowatt-hour cheaper than the electricity cost out of the wall (11.3 ¢ per kilowatt-hour), and it is on par with the LCOE of the peak-shifted utility solar supplied to customers (7.5 ¢ per kilowatt-hour). In 2022 the houses with 100\% battery storage are more expensive than electricity out of the wall and for utility PV with 4 hours of storage delivered to the residence.  In 2030 the houses with 100\% battery storage are less expensive than electricity out of the wall and more expensive than utility PV with 4 hours of storage delivered to the residence.  In 2040 residents can have a net zero home with 100\% battery storage at less cost than utility PV with 4 hours of battery electricity delivered to the home. Even in scenarios where utility companies significantly increase the adoption of utility-scale PV systems and employ batteries to meet the entire electricity demand, it remains economically impractical for Florida residents. The residences can achieve a cost-effective electricity supply by deploying their own PV and battery systems which substantially enhances reliability and resilience, especially during natural disasters such as hurricanes.


\section{Gasoline Equivalent of Residential PV + Battery System LCOE from 2020 to 2050}

\begin{figure}
\centering
\footnotesize
	\includegraphics[width=5in]{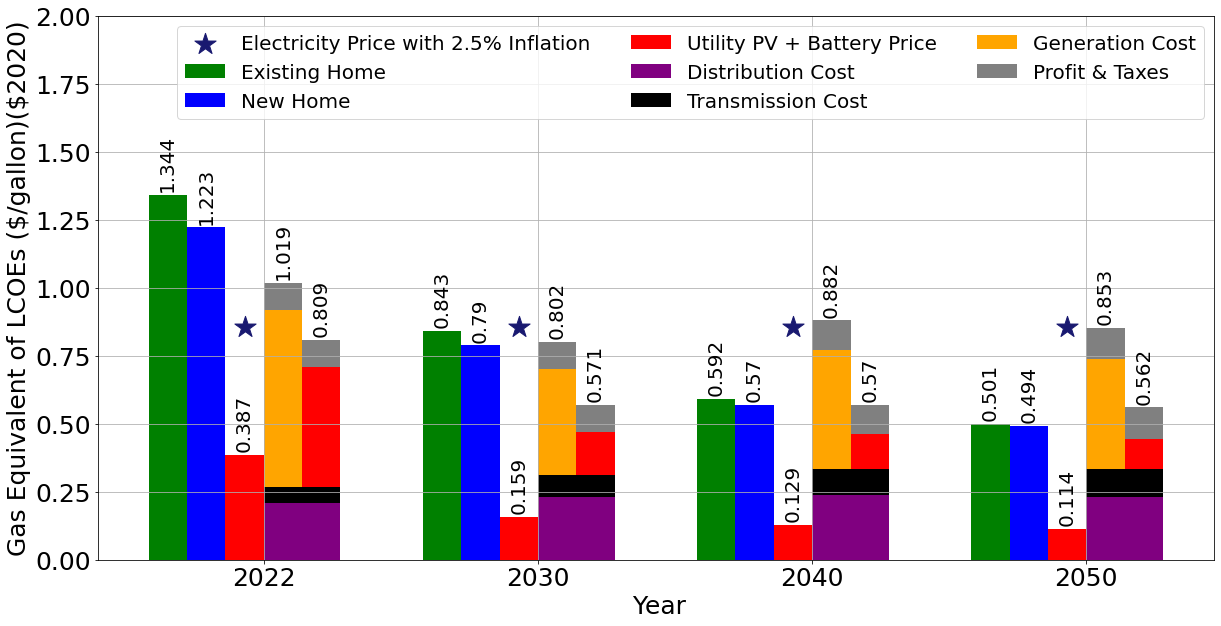}
	\caption{Gas Equivalent of LCOEs (\$/gallon) for PV+Battery Scenario without 30\% PV and Battery ITC.}
    \label{gas_equi_PVBatt}
\end{figure}
The LCOE for rooftop solar in the existing home in 2022 (green column of Figure \ref{LCOE_PV_Batt_Comparison}) is 17.7 ¢ per kWh which is equivalent to \$1.34/gallon (= \$0.177 * 24.2/3.2) which is presented in Figure \ref{gas_equi_PVBatt}.  Since the average price for gasoline in 2023 is \$3.56 per gallon in 2023\$ \cite{GasFLPrice}  or \$3.16 per gallon in 2020\$, the gasoline equivalent of PV on the roof is 42\% the cost of gasoline. 

If the homeowner adds an additional 2.2 kW of PV to the roof to fuel an EV that travels 10,000 miles per year powered on solar electrons the yearly cost is equal to the price of electricity in \$/kWh multiplied by 10,000 miles divided by 3.20 miles/kWh.  For 2022’s rooftop solar electricity price of \$0.058/kWh the yearly EV electricity cost \$181.25.  The 10,000 miles on gasoline costs (10,000/24.2 * \$ 3.16) = \$ 1306 per year.  This leads to a monthly savings of \$93.7 per month for transportation fuel.  In 2022, utilities can produce 100\% renewable electricity at their 75 MW solar fields that have four-hour lithium batteries for EV charging at 5 ¢ per kWh or at the gasoline equivalent of \$0.39 per gallon.  Even if the utilities sold this gasoline-equivalent electricity at \$1.00 per gallon through EV fast chargers they would make far more profit than simply delivering solar electrons over long distances. By 2030 Utility solar fields will all serve as truck stops which will be able to provide 100\% renewable fast charging of electric vehicles and hydrogen fueling of fuel cell vehicles 24 hours a day.

\section{Existing Homes Monthly Electric Bills: Utilities vs. Rooftop PV \& Batteries}\label{MonthlySaving_PVBatt}

\begin{figure}
\centering
\footnotesize
	\includegraphics[width=5.5in]{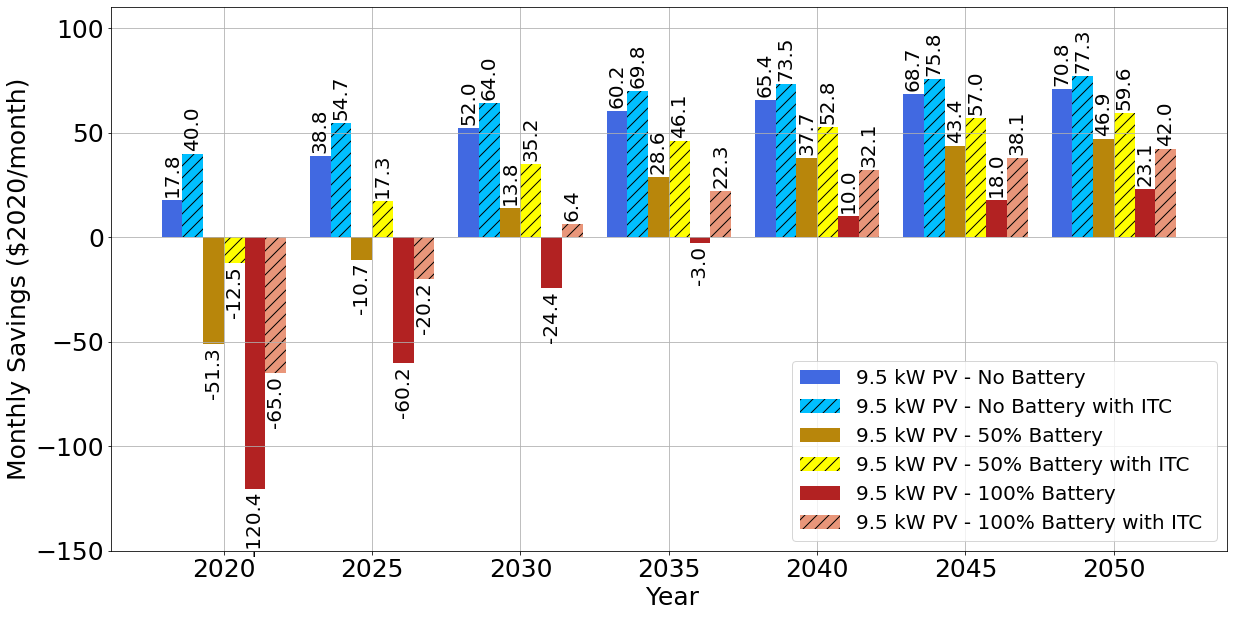}
	\caption{Monthly Savings with PV and Batteries Making the Average Florida Residence Net Zero}
    \label{Monthly_Saving_Battery_ITC}
\end{figure}

Figure \ref{Monthly_Saving_Battery_ITC} displays monthly cost savings from owning rooftop PV with various levels of energy storage (zero, 50\% effective storage, and 100\% effective storage, with and without a 30\% Federal ITC) compared to purchasing electricity from the grid for the years 2020 to 2050. These calculations are based on information from Figures \ref{InstalledCost_PV} and \ref{Installed_BESCost}. As long as net metering is still allowed in Florida, it's not necessary to invest in energy storage to maximize monthly energy savings. In such cases, the blue columns are higher than those with energy storage (50\% and 100\% effective storage, shown in golden, yellow, dark red, and light red). PV on the roof with zero energy storage and without the Federal ITC provides monthly savings in 2020, 2025, 2030, and 2035 of \$18, \$39, \$52, and \$60, respectively. With the Federal ITC, these savings increase to \$40, \$55, \$64, and \$70 per month. Achieving a Net-Zero Energy home with 50\% effective battery storage (21.1 kWh) (golden column) becomes cost-effective before 2030, with savings of \$14 per month. With the Federal ITC, 50\% effective battery storage (21.1 kWh) (diagonal yellow column) becomes cost-effective before 2025, saving \$5 per month. 100\% effective battery storage (red column) becomes cost-effective before 2040, with monthly savings of \$10, allowing 42.2 kWh to not be net-metered. 100\% effective battery storage with the Federal ITC (light red diagonal column) is cost-effective in 2035, saving \$7 per month, and also permitting 42.2 kWh to not be net-metered. Achieving a Net-Zero Energy home with 100\% battery effectiveness allows for the use of all solar electricity to be consumed by the residence, with little to no electricity being net-metered. With the Federal ITC, this peace of mind would cost the residential customer \$98 per month in 2020 and \$44 per month in 2025. So, the crossover point occurs in July 2037 without the ITC and in August 2029 with the ITC!

\section{New Homes Monthly Electric Bills: Utility vs. Rooftop PV, Efficiency Improvements \& Batteries}\label{InstalledCost_PV_efficiencyImprove}

\begin{figure}
\centering
\footnotesize
	\includegraphics[width=5.5in]{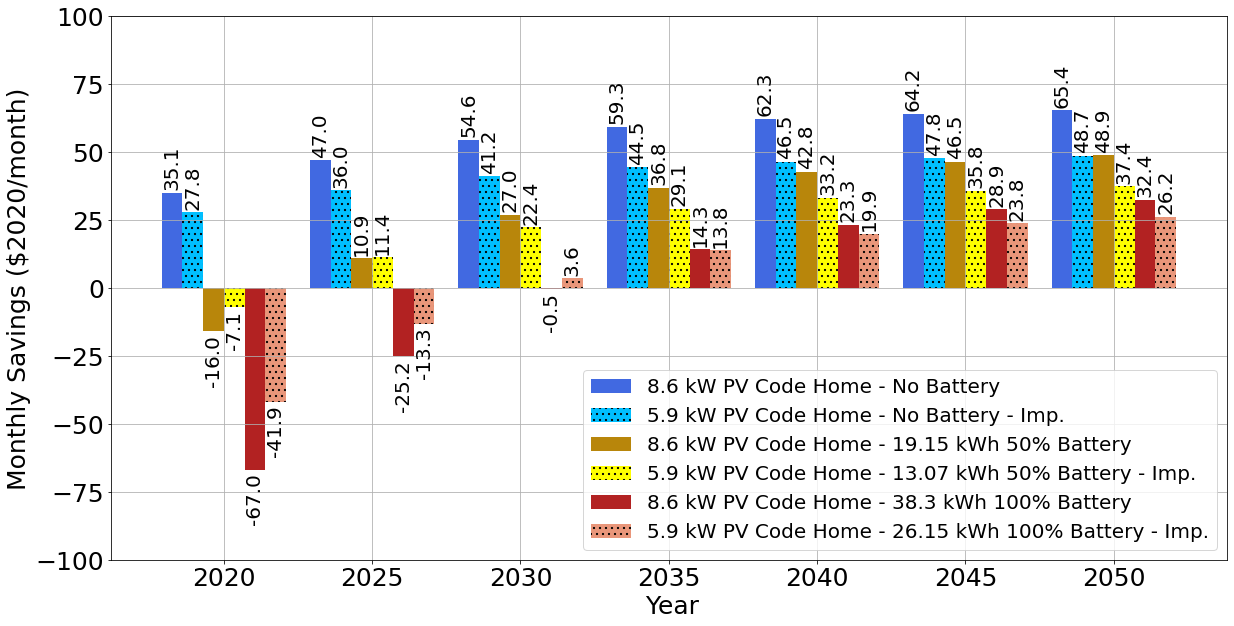}
	\caption{Monthly Savings without the ITC with PV, Batteries and Efficiency Improvements added During the Construction of IECC Home or Improved Home }
    \label{Monthly_Saving_Battery_Improve}
\end{figure}

Table \ref{CodeHomeTable} reveals that the typical Orlando IECC home consumes 12,086 kWh/year in electricity. To transform this home into a Net Zero Electricity Residence, it requires an 8.6 kW solar installation. To achieve complete grid independence on average, more than 33.1 kWh of daily load coverage is needed. Accounting for the expected decline in PV and battery performance over 25 years, 38.3 kWh of storage is required for the Code residence to be consistently grid-independent. Figure \ref{Monthly_Saving_Battery_Improve} illustrates monthly cost savings (without the 30\% Federal ITC) for a new IECC home constructed with rooftop PV and energy storage compared to a new IECC home with 31.7\% cost-effective energy efficiency improvements, rooftop PV, and energy storage for the years 2020, 2025, 2030, 2040, and 2050. The IECC home requires 8.6 kW of PV to be net zero and 19.2 kWh batteries for 50\% effective storage and 38.3 kWh of batteries for 100\% effective storage. The IECC home with 31.7\% energy efficiency improvements requires 5.9 kW of PV and 13.1 kWh of batteries for 50\% effective storage and  26.2 kWh of batteries for 100\% effective storage. While efficiency savings may not be as substantial as adding PV to the existing code home (evident in the difference in heights of blue columns), the more efficient home requires less energy storage, resulting in lower expenses in the earlier years or greater savings up to 2030 compared to the less efficient IECC home to achieve 100\% effective storage. As long as net metering remains permissible in Florida, there is no need to invest in energy storage for the Net-Zero Energy IECC home to maximize monthly energy savings (blue column). In 2020, 2025, and 2030, the savings would be \$35, \$47, and \$55 per month, respectively. The values in Figure \ref{Monthly_Saving_Battery_Improve} without batteries are identical to those in Figure \ref{Monthly_saving_Imp}. To attain a Net-Zero Energy code home with 50\% effective battery storage (19.2 kWh) (golden column), it becomes cost-effective in 2025, resulting in savings of \$11 per month. Achieving 100\% effective battery storage (dark red column) becomes cost-effective shortly after 2030, allowing 38.3 kWh of electricity not to be net-metered. Having 100\% battery effectiveness permits the use of all solar electricity by the residence, with little to no electricity being net-metered. The cost for this peace of mind to withstand utility electricity outages on average would be \$25/month in 2025. Therefore, for the new NET Zero 100\% effective battery code residence, the crossover point occurs in March 2031!\par

As long as net-metering remains permissible in Florida, there would be no need to invest in energy storage for the Net-Zero Energy IECC home with 31.7\% cost-effective improved energy efficiency to maximize monthly energy savings (dotted blue column). In 2020, 2025, and 2030, these savings would amount to \$28, \$36, and \$41 per month, respectively. The values in Figure \ref{Monthly_Saving_Battery_Improve} without batteries remain consistent with those in Figure \ref{Monthly_saving_Imp}. Achieving a Net-Zero IECC home with efficiency improvements and 50\% effective battery storage (19.2 kWh) (dotted yellow column) becomes cost-effective in 2025, resulting in savings of \$11 per month. 100\% effective battery storage (dotted light red column) becomes cost-effective in 2030, saving \$4 per month and allowing 38.3 kWh not to be net-metered. Having 100\% battery effectiveness permits the utilization of all solar electricity by the residence, with little to no electricity being net-metered. The cost of this peace of mind to withstand utility electricity outages for the improved energy efficiency IECC residential customer would be \$13/month in 2025. Therefore, for the new Net Zero 100\% effective battery code residence with energy efficiency improvements, the crossover point occurs in July 2029!

\begin{figure}
\centering
\footnotesize
	\includegraphics[width=5.5in]{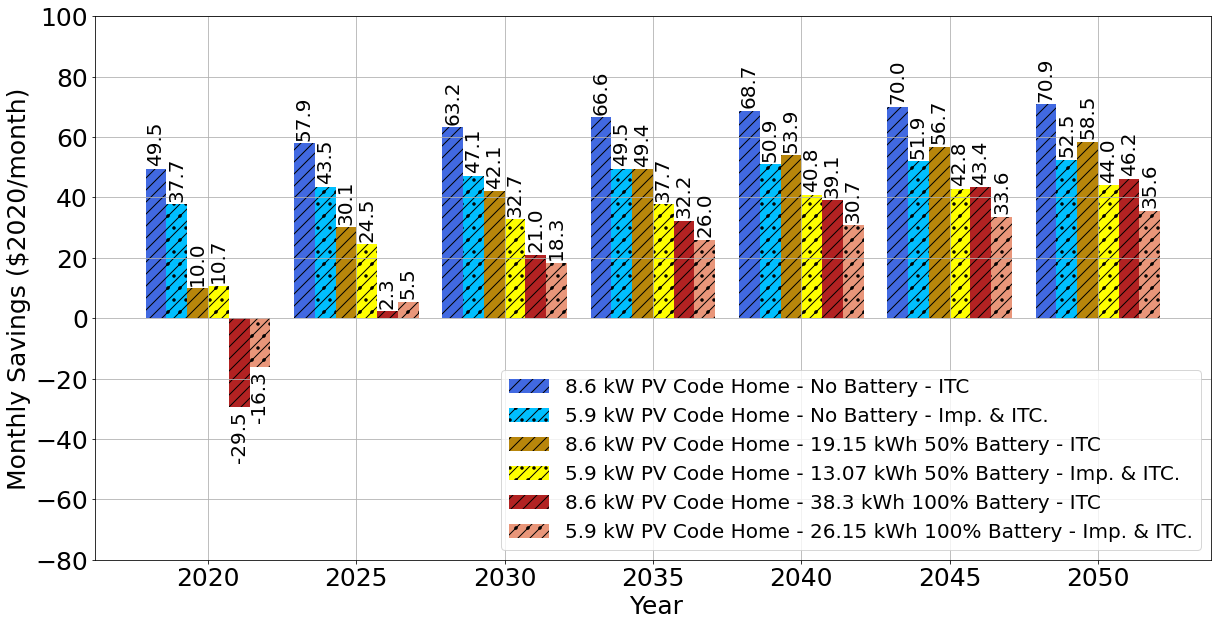}
	\caption{Monthly Savings with the ITC with PV, Batteries, and Efficiency Improvements added before construction of Code Home.}
    \label{Monthly_Saving_Battery_Improve_ITC}    
\end{figure}

Figure \ref{Monthly_Saving_Battery_Improve_ITC} illustrates the monthly cost savings (with the Federal ITC of 30\%) resulting from a new IECC home equipped with PV on the roof and energy storage, compared to a new IECC home with 31.7\% cost-effective energy efficiency improvements, PV on the roof, and energy storage for the years 2020 to 2050. These calculations are based on data from Figures \ref{PVInstalled_residential} and \ref{Installed_BESCost}.

The monthly cost savings for the Net-Zero Energy IECC home (with the Federal ITC of 30\%) with no batteries (blue column) would be \$50, \$58, and \$63 per month in 2020, 2025, and 2030, respectively. The values in Figure \ref{Monthly_Saving_Battery_Improve_ITC} with no batteries are the same as those in Figure \ref{Monthly_saving_Imp}. To achieve Net-Zero Energy status for the IECC home with 50\% effective battery storage (19.2 kWh) (olive column) becomes cost-effective before 2020, resulting in savings of \$13 per month. 100\% effective battery storage (red column) becomes cost-effective in 2025, saving \$9 per month and permitting 38.3 kWh not to be net-metered. Having a Net-Zero Energy home with 100\% battery effectiveness allows the use of all solar electricity to be consumed by the residence, with little to no electricity being net-metered. Therefore, for the new Net-Zero Energy IECC home (with the Federal ITC of 30\%), the crossover point occurred in September 2024.

The monthly cost savings for the Net-Zero Energy IECC home with 31.7\% cost-effective energy efficiency improvements (with the Federal ITC of 30\%) and no batteries (diagonal-dotted light blue column) in 2020, 2025, and 2030 would be \$38, \$44, and \$47 per month, respectively. The values in Figure \ref{Monthly_Saving_Battery_Improve_ITC} with no batteries are the same as those in Figure \ref{Monthly_saving_Imp}. To achieve Net-Zero Energy status for the IECC home with 50\% effective battery storage (19.2 kWh) (diagonal-dotted yellow column) is cost-effective before 2020, resulting in savings of \$11 per month. 100\% effective battery storage (diagonal-dotted pink column) becomes cost-effective before 2025, saving \$6 per month and allowing 38.3 kWh not to be net-metered. Having a Net-Zero Energy home with 100\% battery effectiveness allows the use of all solar electricity to be consumed by the residence, with little to no electricity being net-metered. Therefore, for the new Net-Zero Energy IECC home with 100\% effective battery storage and energy efficiency improvements (with the Federal ITC of 30\%), the crossover point occurs in July 2023.

\begin{table}[]
\centering
\caption{2020 Financial Indexes for Florida Residents with PV + Battery}
\label{PVBatt_Financial_Table2020}
\begin{tabular}{|c|c|c|c|c|c|c|} 
\hline
\textbf{\begin{tabular}[c]{@{}c@{}}Case\\ (PV+ Batt. \\Only)\end{tabular}} &
  \textbf{\begin{tabular}[c]{@{}c@{}}Existing \\ Home\end{tabular}} &
  \textbf{\begin{tabular}[c]{@{}c@{}}New \\ Home\end{tabular}} &
  \textbf{\begin{tabular}[c]{@{}c@{}}New \\ Home\\ with Imp.\end{tabular}} &
  \textbf{\begin{tabular}[c]{@{}c@{}}Existing \\ Home \\ with ITC.\end{tabular}} &
  \textbf{\begin{tabular}[c]{@{}c@{}}New Home \\ with ITC.\end{tabular}} &
  \textbf{\begin{tabular}[c]{@{}c@{}}New Home \\ with ITC \& Imp.\end{tabular}} \\ \hline \hline
\rowcolor[HTML]{EFEFEF} 
\textbf{\begin{tabular}[c]{@{}c@{}}NPV\\ (\$)\end{tabular}}   & -36121  &-28439  &-28955   &-19522  &-14145  &-16773 \\ \hline
\textbf{\begin{tabular}[c]{@{}c@{}}IRR\\ (\%)\end{tabular}}   & -2.8  &-1.93  &-2.2  & -0.67 &0.31  &-0.37 \\ \hline
\rowcolor[HTML]{EFEFEF} 
\textbf{\begin{tabular}[c]{@{}c@{}}SIR\\ \;\end{tabular}}      & 0.49  &0.56   &0.55   & 0.65  &0.74  & 0.7  \\ \hline
\textbf{\begin{tabular}[c]{@{}c@{}}SPB\\ (Year)\end{tabular}} & 15.69 &15.68  &15.21   &10.99   &10.89  &12.09  \\ \hline
\end{tabular}

\end{table}

\begin{table}[]
\centering
\caption{2035 Financial Indexes for Florida Residents with PV + Battery}
\label{PVBatt_Financial_Table2035}
\begin{tabular}{|c|c|c|c|c|c|c|} 
\hline
\textbf{\begin{tabular}[c]{@{}c@{}}Case\\ (PV+ Batt. \\Only)\end{tabular}} &
  \textbf{\begin{tabular}[c]{@{}c@{}}Existing \\ Home\end{tabular}} &
  \textbf{\begin{tabular}[c]{@{}c@{}}New \\ Home\end{tabular}} &
  \textbf{\begin{tabular}[c]{@{}c@{}}New \\ Home\\ with Imp.\end{tabular}} &
  \textbf{\begin{tabular}[c]{@{}c@{}}Existing \\ Home \\ with ITC.\end{tabular}} &
  \textbf{\begin{tabular}[c]{@{}c@{}}New Home \\ with ITC.\end{tabular}} &
  \textbf{\begin{tabular}[c]{@{}c@{}}New Home \\ with ITC \& Imp.\end{tabular}} \\ \hline \hline
\rowcolor[HTML]{EFEFEF} 
\textbf{\begin{tabular}[c]{@{}c@{}}NPV\\ (\$)\end{tabular}}   &-1317   &6849  &-351   &9694  &15409  &7101 \\ \hline
\textbf{\begin{tabular}[c]{@{}c@{}}IRR\\ (\%)\end{tabular}}   & 4.18  &6.53  &4.41  &7.61  &10.52  &6.73\\ \hline
\rowcolor[HTML]{EFEFEF} 
\textbf{\begin{tabular}[c]{@{}c@{}}SIR\\ \;\end{tabular}}      &1.09  &1.37   &1.15   &1.51   &1.88  &1.43  \\ \hline
\textbf{\begin{tabular}[c]{@{}c@{}}SPB\\ (Year)\end{tabular}} &6.77  &5.69  &8.47   &4.74   &3.98  &7.31  \\ \hline
\end{tabular}

\end{table}


Table \ref{PVBatt_Financial_Table2020} shows the 2020 financial decision indexes for Florida homes with PV and battery. Six different scenarios were considered again to have a better comparison; Florida residents who equip their “existing home” with PV and battery after the construction;  PV and battery installed during the construction of a “new home”; PV and battery and efficiency improvements installed during the construction of a “new home with efficiency improvements”; and each of these three homes with 30\% PV and 30\% battery “ITC”s. In 2020 the NPV is negative and the SIR is less than 1.0 for all cases.  This shows that 100\% batteries are not a good investment in 2020. The range of SPB is 10.89 to 15.69 years which would yield monthly savings with a 15 or 30-year mortgage.  Table IV shows the 2035 financial decision indexes for Florida homes with PV and battery.  In 2035 the NPVs are mostly positive and the SIR is greater than 1.0 for all cases. The range of SPB is 3.98 to 8.47  years which would yield monthly savings with a 10, 15, or 30-year mortgage.  

The next question to answer is, what year will Florida residents pay less per month on their energy needs for their residences and their fuel for transportation by adding a single EV, while still having reliable and automatic backup power (V2H) when the electric grid is not available? 

\section{Existing Homes Monthly Electric Bills: Utility vs. Rooftop PV, Electric Vehicle to Home (for Backup Power)}\label{V2G}

\begin{figure}
\centering
\footnotesize
	\includegraphics[width=5.5in]{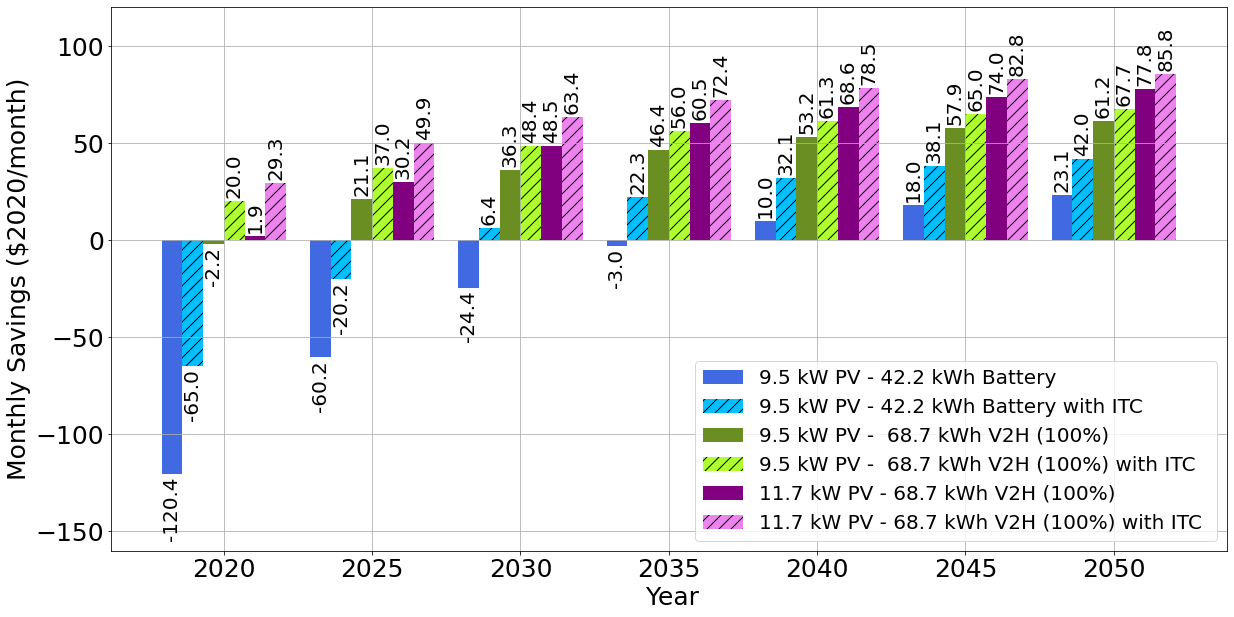}
	\caption{Monthly Savings with PV, Batteries, and V2H with and without ITC Making the Average Florida Residence Net Zero}
    \label{V2G_IncomeTax}
\end{figure}

Figure \ref{Monthly_Saving_Battery_ITC} illustrated the monthly cost savings resulting from owning 9.5 kW of PV on the roof with varying levels of energy storage: zero energy storage, 50\% effective storage (21.1 kWh), and 100\% effective storage (42.2 kWh). These scenarios are presented with and without a Federal Investment Tax Credit (ITC) of 30\%. The comparison is made against purchasing electricity from the grid. In Figure \ref{V2G_IncomeTax}, two different setups are presented. The first involves 9.5 kW of PV on the roof with 100\% effective storage (42.2 kWh) (represented by dark and light blue columns). The second scenario features 9.5 kW of PV on the roof, a bidirectional charger (\$ 6000 installed cost), and a Vehicle-to-Home (V2H) capable electric car (represented by dark and light green columns). Additionally, the setup includes 11.7 kW of PV on the roof (represented by dark and light purple columns), a bidirectional charger, and a V2H-capable electric car. The upfront cost of 100\% batteries on the wall proves to be cost-effective for backup power before 2030 with the ITC. 

Considering the relatively low upfront cost of the bidirectional charger (\$6,000) compared to wall batteries, all of the homes in Figure \ref{V2G_IncomeTax} can cost-effectively provide backup power in 2021 using existing PV on the roof and the homeowner’s EV  well before the bi-directional chargers and EVs are ready for V2H in 2024. As the average electric vehicle available today has a range of 220 miles and has a 68.7 kWh battery the efficiency is then 3.20 miles/kWh.  The addition of 2.2 kW of PV to the 9.5 kW home with V2H capability provides extra electricity, enabling an additional 3,080 kWh/yr or approximately 10,000 miles/year of driving powered by solar electricity.  The 10,000 miles powered on solar electrons costs (\$ 0.04 x 3123)= \$ 124.92 per year.  The 10,000 miles on gasoline costs (10,000 / 24.2 x \$3.16) = \$ 1305.8 per year.  The \$ 1180.9 per year in savings on transportation fuel is an additional \$ 98.4 monthly savings

\begin{figure}
\centering
\footnotesize
	\includegraphics[width=5.5in]{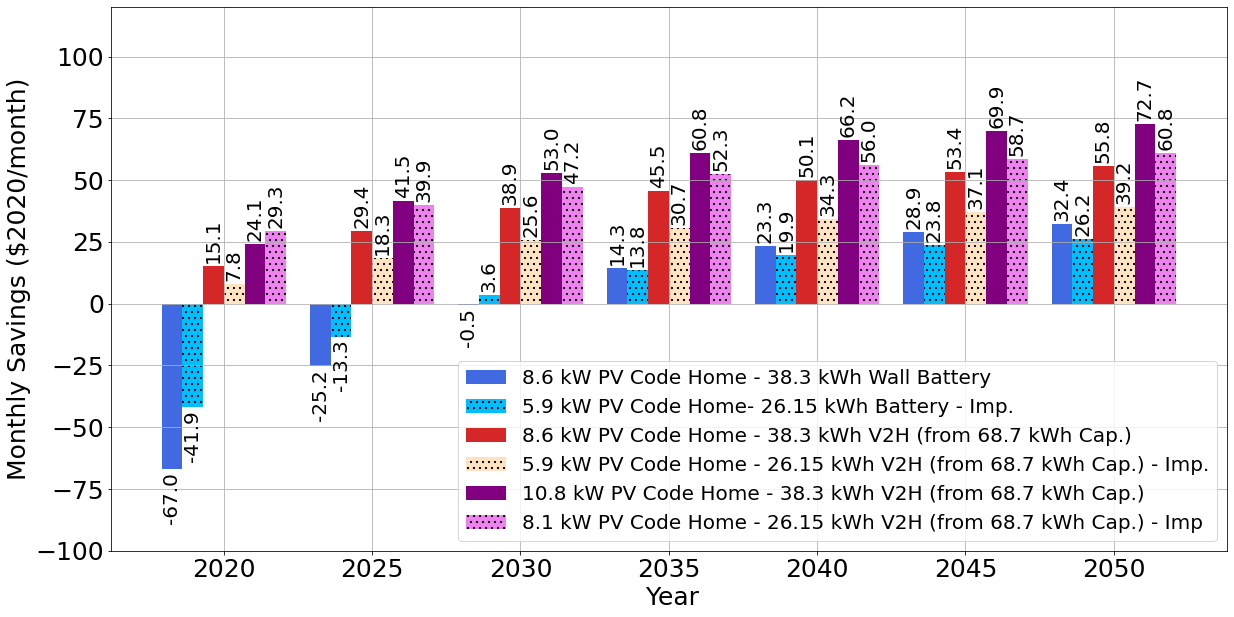}
	\caption{Monthly Savings with PV, Batteries, and V2H with Efficiency Improvement Making the Average Florida Residence Net Zero}
    \label{V2G_improvement}
\end{figure}
Figure \ref{V2G_improvement} follows the same style as Figure \ref{Monthly_Saving_Battery_ITC}. However, it includes IECC Code Homes with and without efficiency improvements, PV, batteries, and bidirectional chargers added during construction. The first column (dark blue) represents the code home without efficiency improvements, featuring 8.6 kW of PV on the roof and 100\% effective storage (38.3 kWh). The second column (light blue) represents the code home with efficiency improvements, equipped with 5.9 kW of PV on the roof and 100\% effective storage (26.2 kWh). The third column (dark red) represents the code home without efficiency improvements, with 8.6 kW of PV on the roof, and 100\% effective storage (38.3 kWh) provided by a bidirectional charger and a V2H-capable electric car (with 68.7 kWh capacity). The fourth column (light red) represents the code home with efficiency improvements, equipped with 5.9 kW of PV on the roof, and 100\% effective storage (26.2 kWh) provided by a bidirectional charger and a V2H-capable electric car (with 68.7 kWh capacity). The fifth column (purple) represents the code home without efficiency improvements, featuring 10.8 kW of PV on the roof, and 100\% effective storage (38.3 kWh) provided by a bidirectional charger and a V2H-capable electric car (with 68.7 kWh capacity). The sixth column (light purple) represents the code home with efficiency improvements, equipped with 8.1 kW of PV on the roof, and 100\% effective storage (26.2 kWh) provided by a bidirectional charger and a V2H-capable electric car (with 68.7 kWh capacity). Considering the relatively low upfront cost of the bidirectional charger (\$6,000) compared to wall batteries, all of the homes in Figure 17 can cost-effectively provide backup power in 2020 using existing PV on the roof and the homeowner's EV. In Figure \ref{LCOE_PV_Residential1}, the LCOE for rooftop solar in the IECC home with the ITC in \$ 2020 stands at 2.9 ¢ per kWh or 0.91 ¢ per mile. When compared to an average gasoline vehicle getting 24.2 mpg, this is equivalent to paying \$ 0.22 per gallon of gasoline (see Figure \ref{gas_equi_PVonly_ITC}), which is approximately one-fourteenth of the gasoline cost in Florida (\$ 3.16/gal).   The 10,000 miles powered on solar electrons costs (\$ 0.029 x 3123)= \$ 90.56 per year.  The 10,000 miles on gasoline costs (10,000 / 24.2 x \$3.16) = \$ 1305.8 per year.  The \$ 1215.2 per year in savings on transportation fuel is an additional \$ 101.3 monthly savings.

\begin{figure}
\centering
\footnotesize
	\includegraphics[width=5.5in]{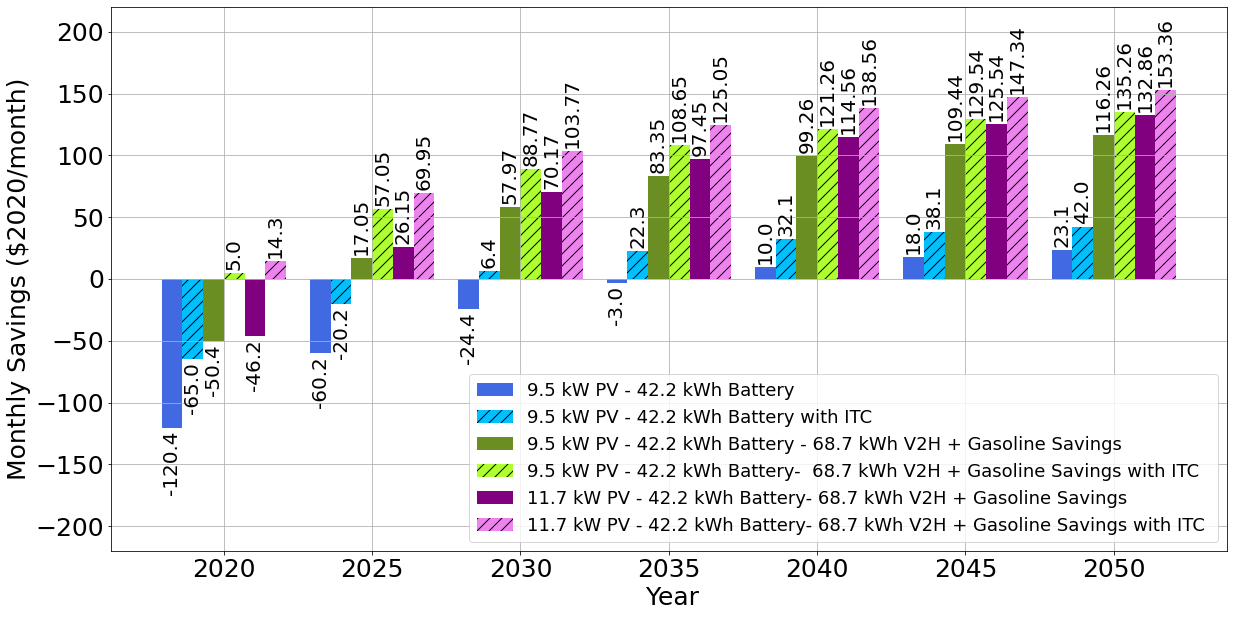}
	\caption{Monthly Savings with adding PV, Batteries, and V2H Capability and Gasoline Savings with and without ITC Making the Average Florida Residence Net Zero}
    \label{V2H_transportationsaving_extraPV}
\end{figure} 

Figure \ref{V2H_transportationsaving_extraPV} shows the monthly savings of an existing home by adding 2.2 kW extra PV (9.5 kW + 2.2 kW = 11.7 kW) for EV charging, a 42.2 kWh battery, and V2H capability with EV battery capacity of 68.7 kWh with and without considering the ITC. The difference between Figure \ref{V2H_transportationsaving_extraPV} and Figure \ref{V2G_IncomeTax} is that savings from gasoline price is added to the last four case studies which clearly indicates that even in the year 2020, Florida residents can save money by having an 11.7 kW PV, a 42.2 kWh battery and average EV with 68.7 kWh battery capacity as well as using PV and battery ITC.   

The answer to the question, "What year will Florida residents pay less per month for their energy needs in their residences and their fuel for transportation by adding a single EV, while still having reliable and automatic backup power (V2H) when the electric grid is not available?" is that the cost of the installed bidirectional charger is low enough that V2H-capable EVs can cost-effectively provide backup power when the grid is out for net zero energy existing homes in 2025 and were cost-effective for net zero energy new homes in 2020. The rate-limiting step to providing new homes with EV backup power is the date when V2H capability and bidirectional chargers are readily available.

\section{Impact of the NZEB on Future H2 Production Cost }\label{H2Cost_}

\begin{figure}
\centering
\footnotesize
	\includegraphics[width=5.5in]{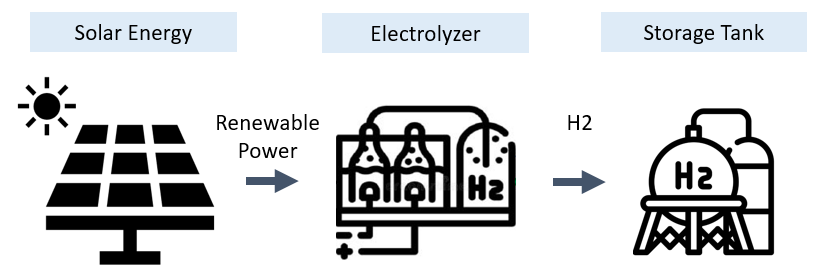}
	\caption{Renewable Electrolysis for H2 Production and Storage at Utility Scale.}
    \label{H2_design}
\end{figure}

\begin{figure}
\centering
\footnotesize
	\includegraphics[width=5.8in]{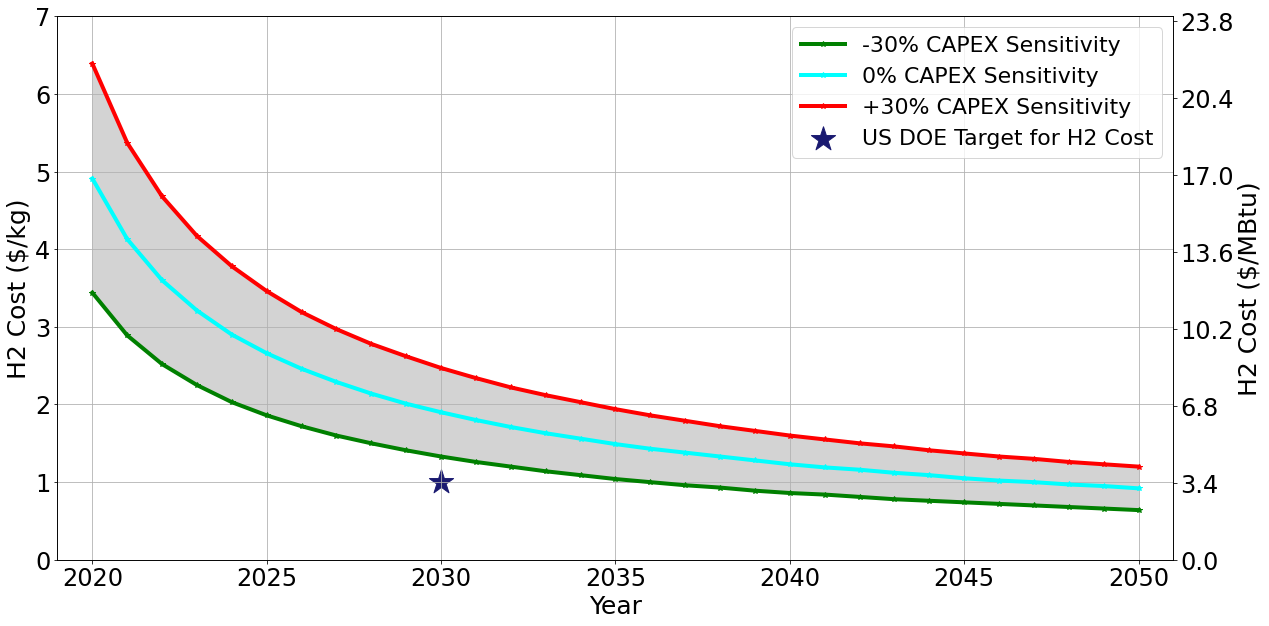}
	\caption{H2 Cost Forecasts in (\$/kg) and (\$/MBtu)}
    \label{H2Cost}
\end{figure}

As mentioned earlier, with lower monthly payments and the option of backup power through rooftop PV and battery storage systems, NZEBs can indirectly support the power grid. For example, supplying buildings' electricity demand with rooftop PV and batteries can significantly enhance voltage profiles \cite{haggi2021risk} and increase the resilience and reliability of the system \cite{haggi2022proactive}\cite{haggi2021proactive}. This enables the creation of virtual power plants and reduces the amount of time utilities need to operate their most expensive power plants. The authors anticipate a substantial increase in the demand for hydrogen electrolysis energy by 2030, surpassing the levels shown in Figure \ref{FLA_100_baseUpdate}, due to the cost-effectiveness of NZEBs. By 2030, it is conceivable that utilities will allocate a significant portion of their solar-generated electricity for hydrogen production through electrolysis, as illustrated in Figure \ref{H2_design}. This approach may prove more economically advantageous than transmitting electricity over long distances to residential and commercial buildings. NZEBs can aid utilities by reducing the overall residential load through energy management systems and facilitating energy exchange with neighbors, a concept known as peer-to-peer energy sharing among residential buildings \cite{haggi2023p2p,haggi2021multi}. Consequently, this leads to an increase in the availability of utility-scale renewables, subsequently driving down hydrogen production costs for both the power and transportation sectors while simultaneously reducing carbon emissions. Figure \ref{H2Cost} presents hydrogen cost forecasts, factoring in a 30\% sensitivity analysis in PV Levelized Cost of Electricity (LCOE), electrolyzer costs, and storage tank Capital Expenditure (CAPEX) costs for PV + Electrolyzer + Storage tank systems with optimal sizing (see  Figures \ref{PV_LCOEE}, Figure \ref{ELX_CAPEX}, and Figure \ref{Tank_CAPEX}, in the Appendix sections for more discussion). With the widespread adoption of NZEBs, utilities can produce hydrogen at reduced costs, ultimately contributing to the realization of the U.S. Department of Energy's (DOE) net-zero emission targets by the year 2050 \cite{haggi2022hydrogen}. In conclusion, NZEBs have the potential to revolutionize the energy landscape by reducing energy consumption, enhancing grid performance, and making hydrogen production more economically viable. These developments align with national and global sustainability goals, ultimately contributing to a cleaner and more efficient energy future.


\section{Conclusion}

This study delved into the pivotal role of NZEBs within the context of Florida's ongoing energy transition. A comprehensive techno-economic analysis used various financial indexes to examine the savings of incorporating a 9.5 kW PV system into existing Florida residences. The implications of introducing a 42 kWh battery storage system that on average makes the home grid-independent but also provides homeowners with reliable backup power during grid disruptions. There are significant financial advantages to integrating PV panels, energy efficiency improvements, and batteries during the construction phase of new homes. Installation at the time of construction is more cost-effective than retrofitting an existing home.  For the average  Florida residence adding 9.5 kW PV and a 42.2 kWh battery (with the Federal ITC) a positive monthly savings over grid electricity occurs in August of 2029.   On the other hand, for a newly built IECC Orlando home with 8.6 kW PV and
38.3 kWh battery (with the Federal ITC), and with energy efficiency improvements the crossover point was July 2023. The addition of an EV to the home can provide additional backup power through V2H and substantial monthly savings (~\$100) by replacing gasoline costs. Today, Florida residents can save on their monthly costs of electricity and gasoline if they have both PV on their roof and a battery system (sized for average daily residence load), and benefit from the federal ITC.  NZEBs also have the potential to increase grid stability by reducing overall electricity consumption, incentivizing customers to participate in peer-to-peer energy-sharing markets, and increasing available capacity for utility-scale renewable energy generation for transportation applications instead of supplying residential electricity demand. It is more cost-effective for the utilities to operate utility-scale PV to be used for on-site EV fast-charging and hydrogen production through electrolysis for fuel cell vehicles or blending with natural gas, minimizing the costs for transmission and distribution. Consequently, all these advancements contribute to a reduction in the cost of H2 production—a pivotal step toward achieving the ambitious decarbonization targets set by the United States for 2030 and 2050. In essence, this study highlights the multifaceted benefits of NZEBs in the context of Florida’s energy landscape.


\break

\section{Appendix}

\begin{figure}[h]
\centering
\footnotesize
	\includegraphics[width=5.25in]{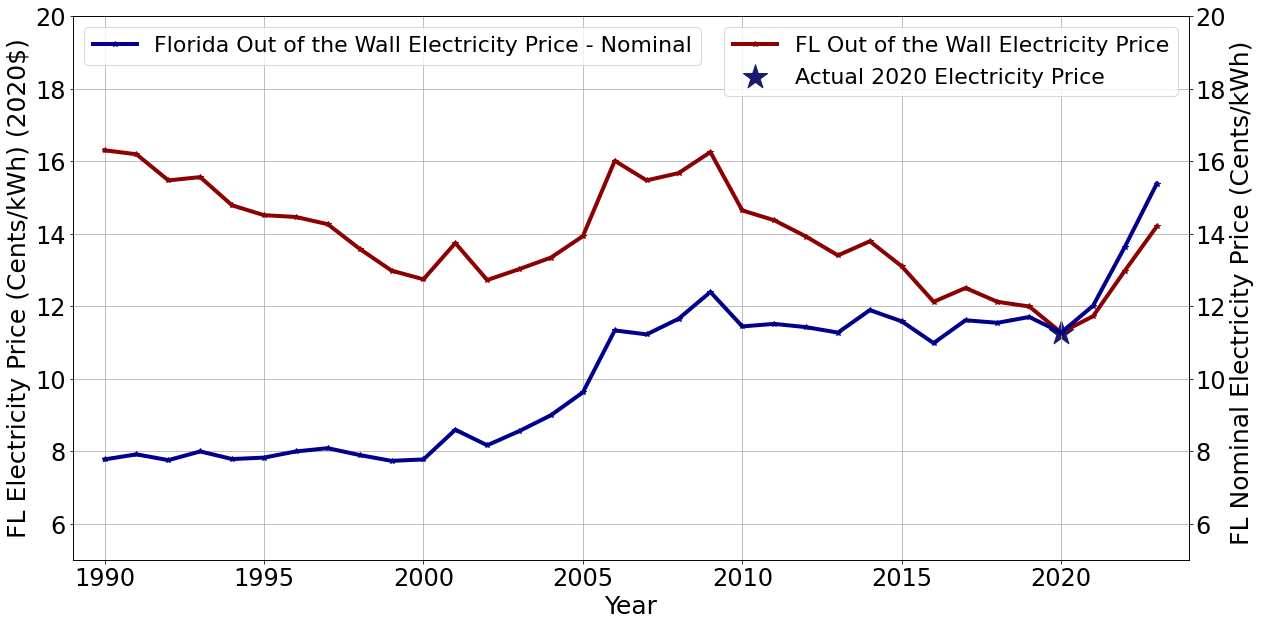}
	\caption{Florida Historical “out of the wall,” Electricity Prices in \$2020.}
    \label{Historic_prices}
\end{figure}

Figure \ref{Historic_prices} shows the historical "out of the wall" residential electricity prices for the state of Florida in \$2020 \cite{EIAprice}. In 2020 the cost of residential electricity out of the wall was 11.3 ¢ per kWh which is the lowest cost for residential customers in 2020\$ for 1990 to 2023.  The “star” represents the price of 11.3 ¢ per kWh in \$2020 used for the calculation of the monthly savings.  For all calculations in future years the 2020 cost of electricity out of the wall, 11.3 ¢ per kWh in 2020\$ does not change with time as the price is assumed to increase at the general rate of inflation, 2.5\%.  This is a very conservative value of electricity out of the wall compared to the cost of electricity out of the wall in 2023 which is 15.36 ¢ per kWh in 2023\$ or 14.3 ¢ per kWh in 2020\$.


\break

\begin{figure}[h]
\centering
\footnotesize
	\includegraphics[width=5.2in]{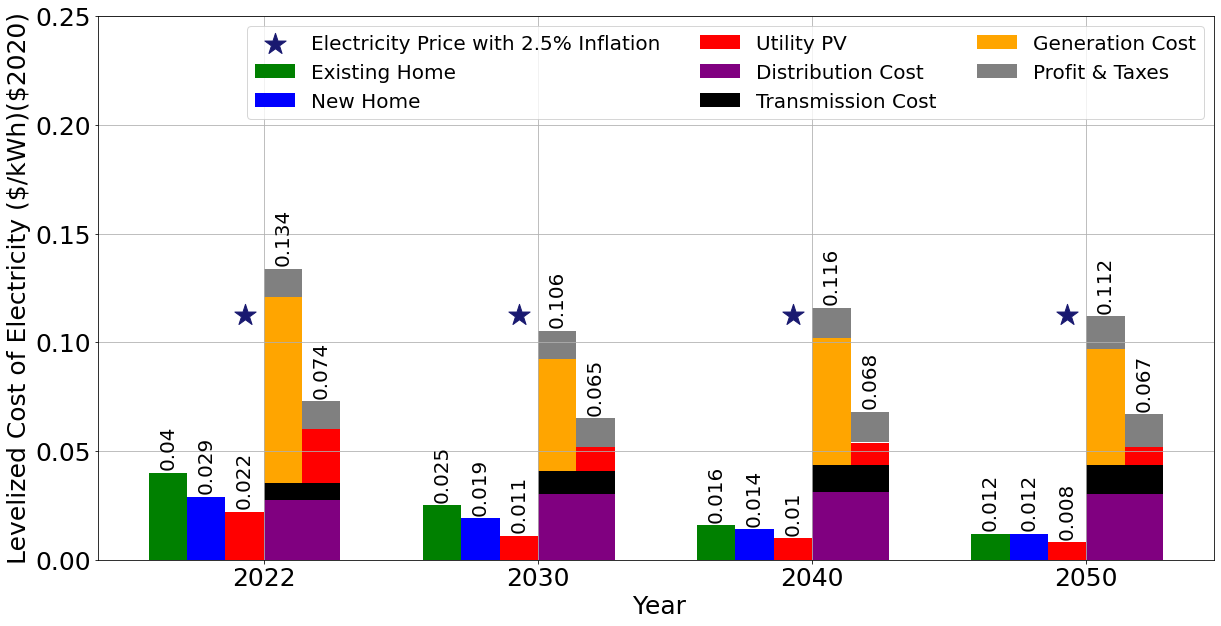}
	\caption{Florida Levelized Cost of Electricity; Two Residential PV Homes, Utility PV, and Residential Electricity “out of the wall,” with 30\% Federal Income Tax Credit Applied to Residential PV}
    \label{LCOE_PV_Residential1}
\end{figure}

\begin{figure}[h]
\centering
\footnotesize
	\includegraphics[width=5.2in]{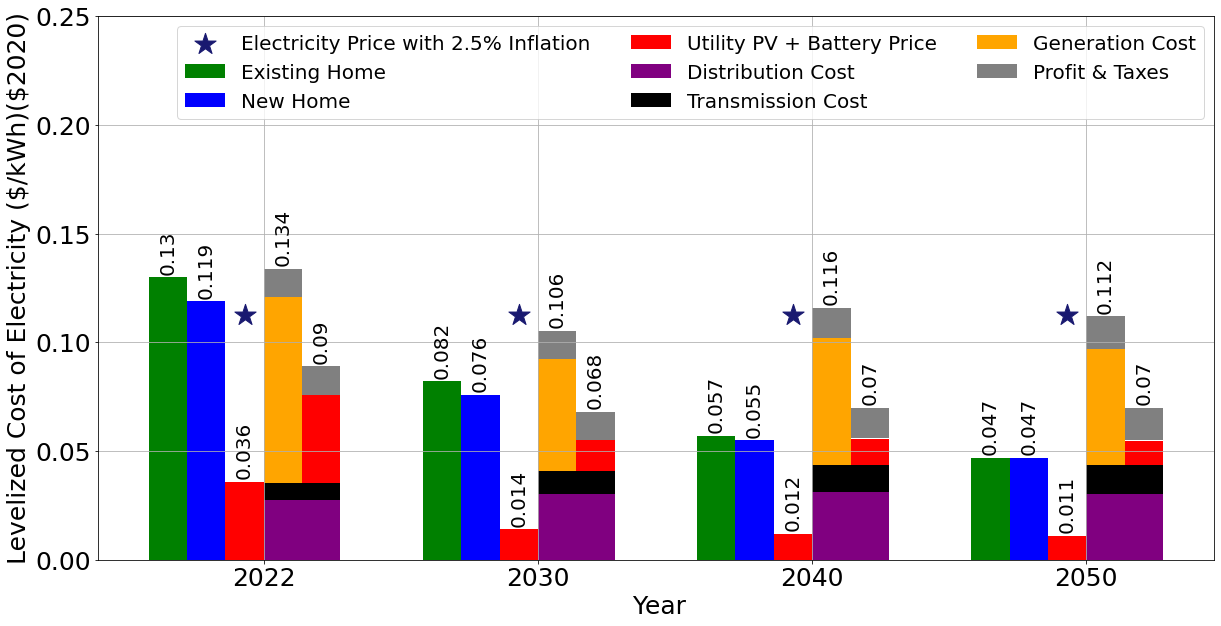}
	\caption{Florida Residential PV + Battery (100\% effective storage) and Utility PV + Battery (4-hour) "out of the wall" Cost of Electricity, with the 30\% Federal ITC for PV and Battery}
    \label{LCOE_PV_Batt_Comparison1}
\end{figure}

Figure \ref{LCOE_PV_Residential1} and Figure \ref{LCOE_PV_Batt_Comparison1} illustrate the LCOE of standalone PV and PV integrated with Battery systems with a 30\% federal ITC, respectively. In comparison to Figure \ref{LCOE_PV_Residential} and Figure \ref{LCOE_PV_Batt_Comparison}, the LCOE figures for both existing and new residential properties in Florida are notably reduced, indicating significant potential cost savings for residents.

\begin{figure}[h]
\centering
\footnotesize
	\includegraphics[width=3.75in,height=1.66in]{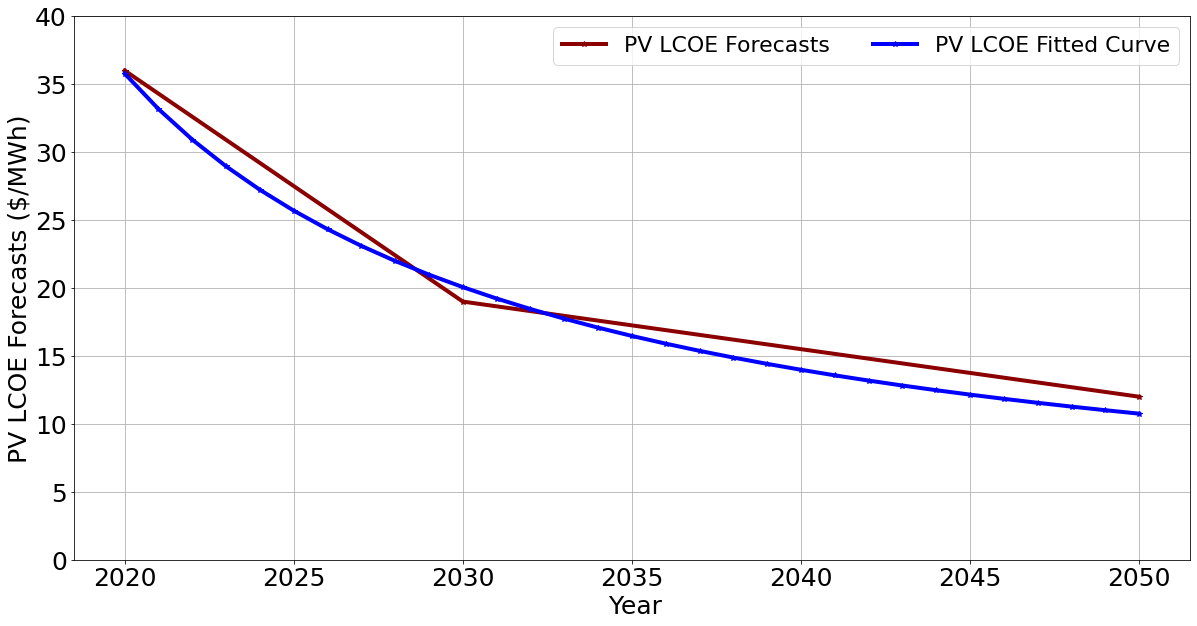}
	\caption{PV LCOE Forecasts}
    \label{PV_LCOEE}
\end{figure}

\begin{figure}[h]
\centering
\footnotesize
	\includegraphics[width=3.75in,height=1.66in]{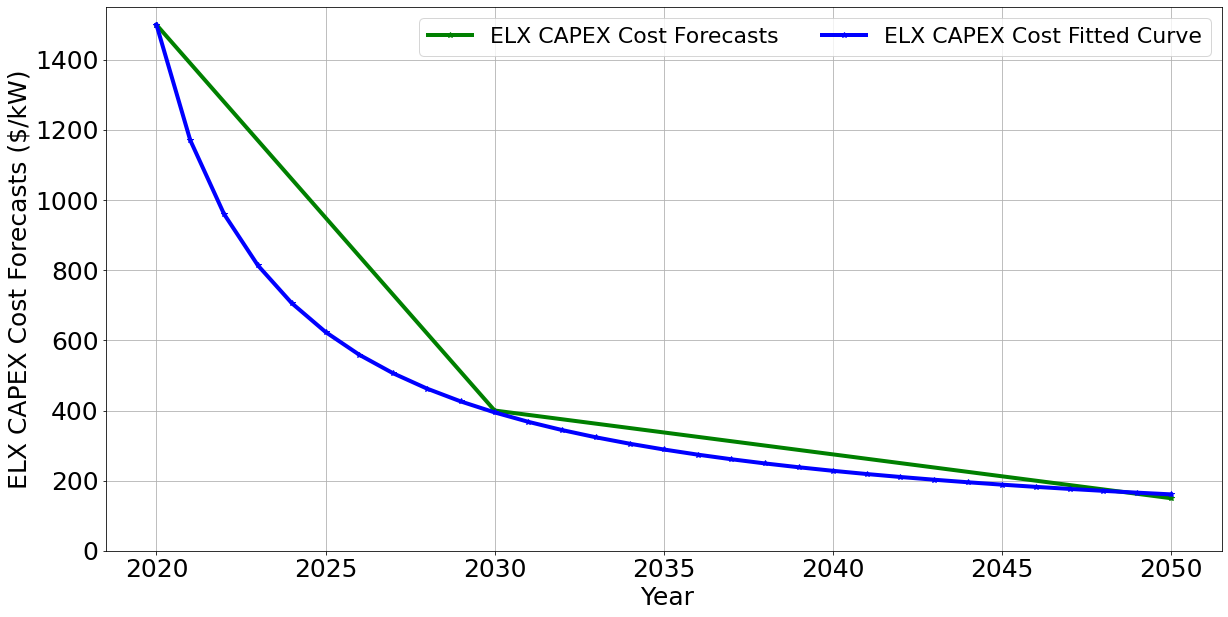}
	\caption{Electrolyzer CAPEX Forecasts}
    \label{ELX_CAPEX}
\end{figure}

\begin{figure}[h]
\centering
\footnotesize
	\includegraphics[width=3.75in,height=1.66in]{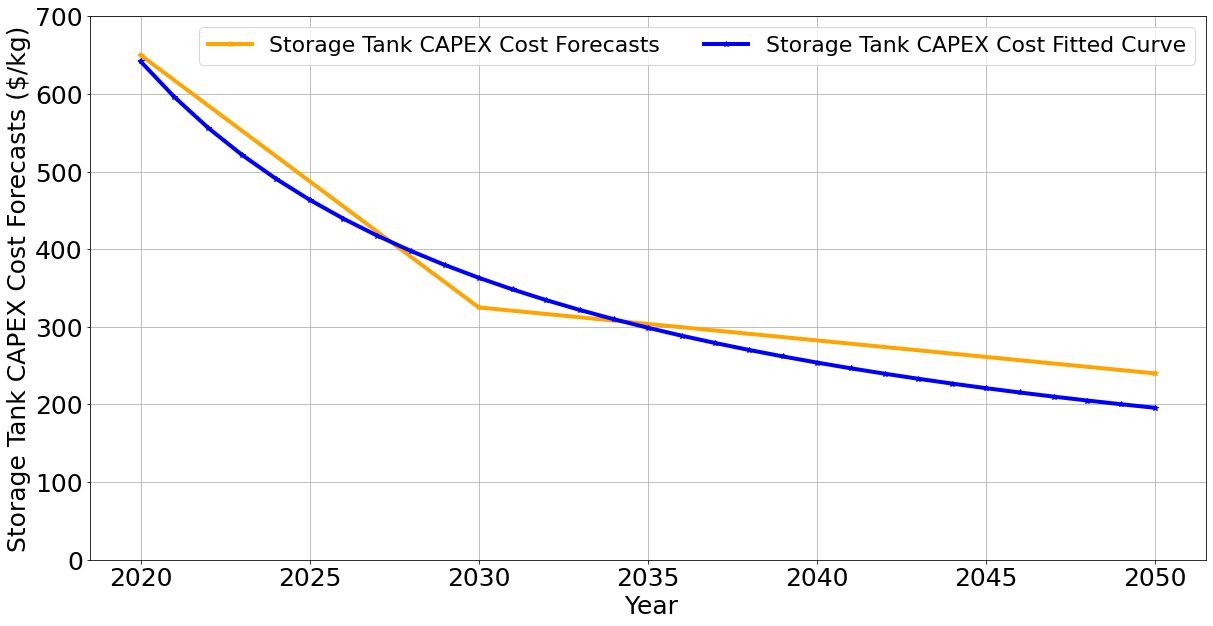}
	\caption{Storage Tank CAPEX Forecasts}
    \label{Tank_CAPEX}
\end{figure}

\break
Cost-related parameters forecast from 2020 to 2050 as well as their fitted curve for the PV LCOE (can be found in \cite{NRELATB}), Electrolyzer, and Storage tank CAPEX costs (can be found in \cite{ELXCapexDOE} and \cite{TankCapexDOE}, respectively) are shown in Figure \ref{PV_LCOEE}, Figure \ref{ELX_CAPEX}, and Figures \ref{Tank_CAPEX}, respectively.


\break

\begin{figure}[h]
\centering
\footnotesize
	\includegraphics[width=5in]{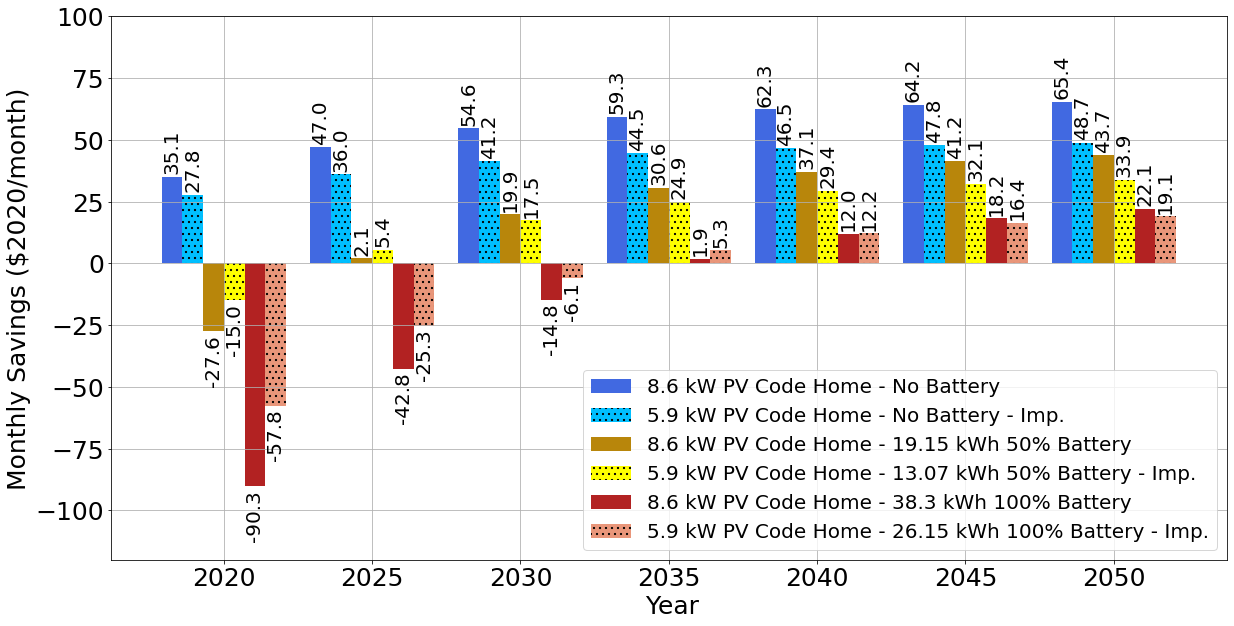}
	\caption{Monthly Savings without the ITC with PV, Batteries and Efficiency Improvements added after the Construction of IECC Home or Improved Home.}
    \label{Fig23}
\end{figure}

\begin{figure}[h]
\centering
\footnotesize
	\includegraphics[width=5in]{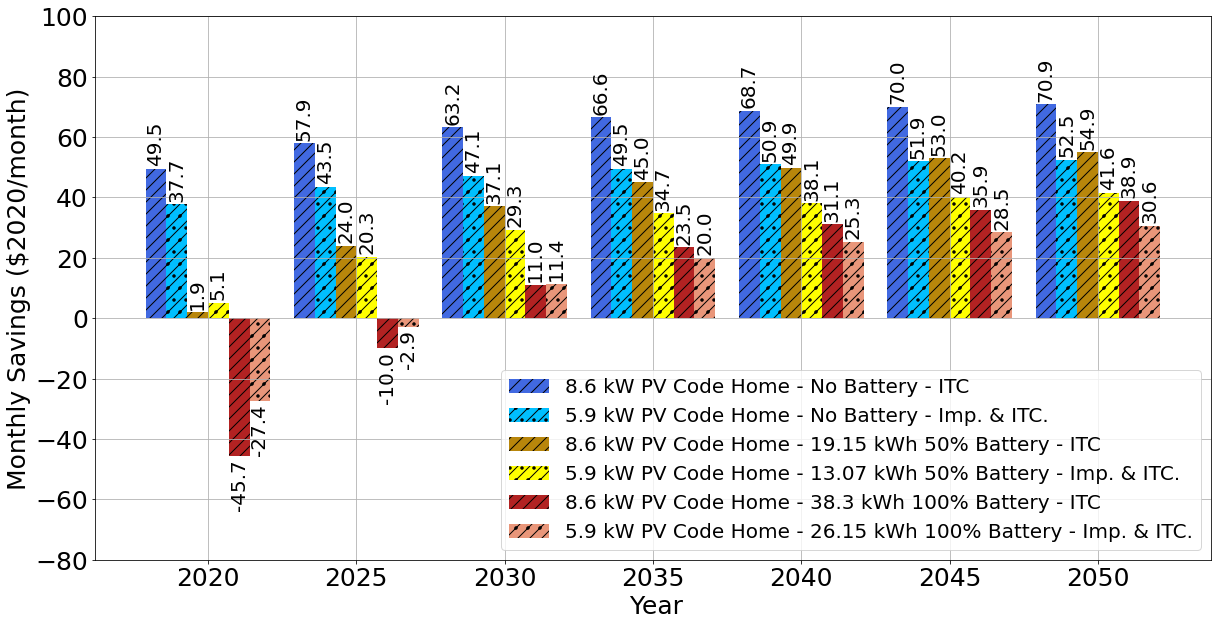}
	\caption{Monthly Savings with the ITC with PV, Batteries, and Efficiency Improvements added after the construction of Code Home.}
    \label{Fig24}
\end{figure}

Figures \ref{Fig23} and \ref{Fig24} show monthly savings with PV, batteries, and efficiency improvements added after the construction of the IECC Home with and without ITC. 

\break

\begin{figure}[h]
\centering
\footnotesize
	\includegraphics[width=5in]{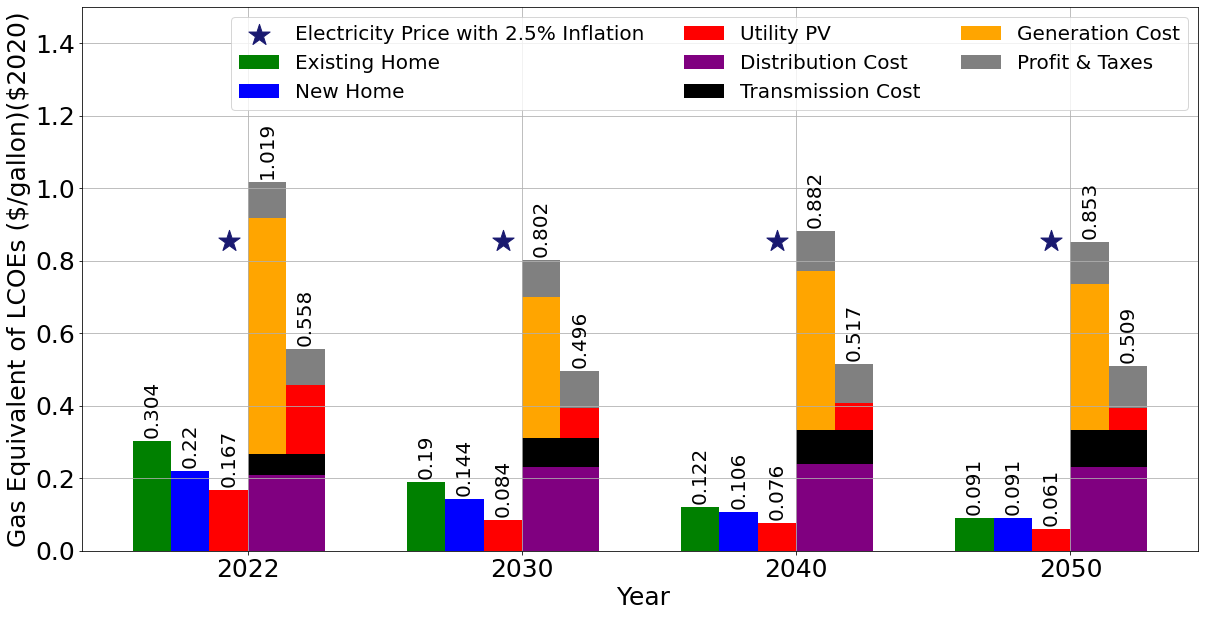}
	\caption{Gas Equivalent of LCOEs (\$/gallon) for PV Only Scenario with 30\% PV ITC.}
    \label{gas_equi_PVonly_ITC}
\end{figure}

\begin{figure}[h]
\centering
\footnotesize
	\includegraphics[width=5in]{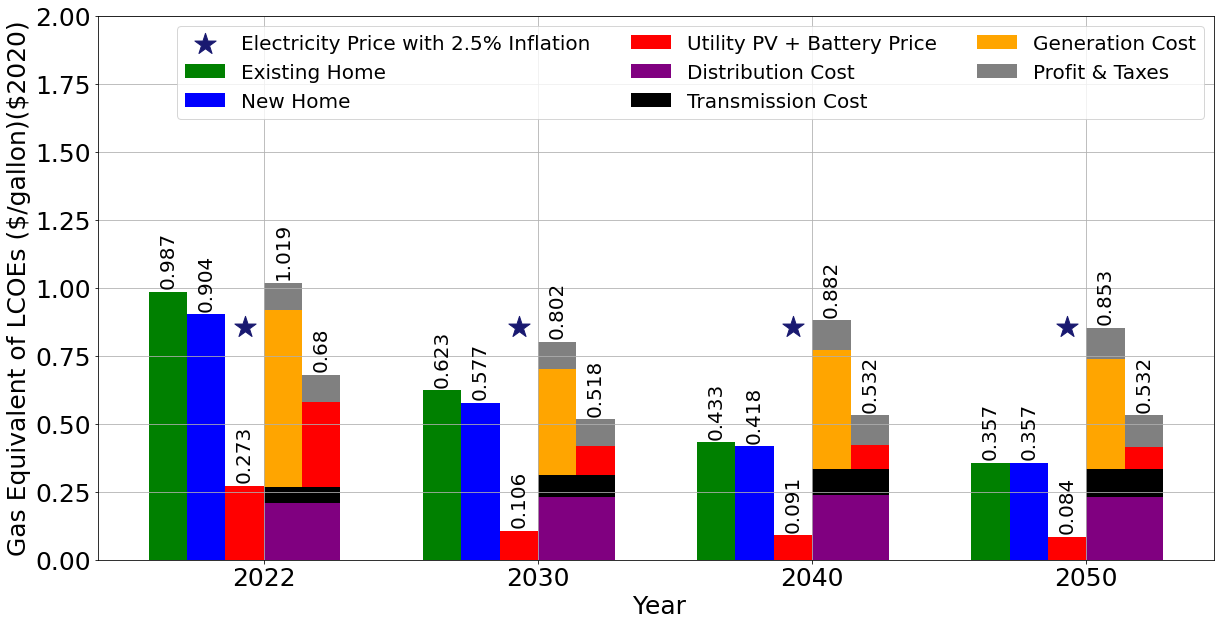}
	\caption{Gas Equivalent of LCOEs (\$/gallon) for PV+Battery Scenario with 30\% PV and Battery ITC.}
    \label{gas_equi_PVBatt_ITC}
\end{figure}

The following figures show the gas equivalent of LCOEs in (\$/gallon); 1) PV-only scenario without and with 30\% federal ITC (presented in Figure \ref{gas_equi_PVonly} and Figure \ref{gas_equi_PVonly_ITC}, respectively), 2) PV + Battery scenario without and with 30\% federal ITC (presented in Figure \ref{gas_equi_PVBatt} and Figure \ref{gas_equi_PVBatt_ITC}, respectively). 

Figure \ref{LCOE_PV_Residential1} showed that the LCOE for rooftop solar in the existing home with the ITC in \$2020 stands at 4 ¢ per kWh or 1.25 ¢ per mile. When compared to an average gasoline car getting 24.2 mpg \cite{USDOEEERE}, this is equivalent to paying \$ 0.304 per gallon of gasoline (see Figure \ref{gas_equi_PVonly_ITC}), which is approximately one-ninth of the gasoline cost in Florida today (\$ 3.40/gal).   The 10,000 miles powered on solar electrons costs (\$ 0.04 x 3123)= \$ 124.92 per year.  The 10,000 miles on gasoline costs (10,000 / 24.2 x \$3.40) = \$ 1404.9 per year.  The \$ 1280.7 per year in savings on transportation fuel is an additional \$ 106.7 monthly savings.


\break

The following table shows the important parameters that were used in the simulation. 

\begin{table}[h]
\centering
\begin{tabular}{|cc|}
\hline
\multicolumn{2}{|c|}{\cellcolor[HTML]{EFEFEF}\textbf{All Inputs Used in This Paper}} \\ \hline
\multicolumn{1}{|c|}{Average Existing Home PV (kW)}                & 9.5    \\ \hline
\multicolumn{1}{|c|}{IECC Home PV (kW)}                   & 8.6    \\ \hline
\multicolumn{1}{|c|}{IECC Home PV with Improvement (kW)}  & 5.6    \\ \hline
\multicolumn{1}{|c|}{PV Inverter Cost (\$/pWdc)}          & 0.1    \\ \hline
\multicolumn{1}{|c|}{PV Inverter Life (Years)}            & 15     \\ \hline
\multicolumn{1}{|c|}{PV Degradation Factor (\% per year)} & 0.5    \\ \hline
\multicolumn{1}{|c|}{100\% Battery Capacity (kWh)}        & 42.21  \\ \hline
\multicolumn{1}{|c|}{100\% Battery Capacity for Code Home (kWh)}            & 38.3   \\ \hline
\multicolumn{1}{|c|}{100\% Battery Capacity for Improved Code Home (kWh)}   & 26.15  \\ \hline
\multicolumn{1}{|c|}{Battery efficiency (\%)}             & 95     \\ \hline
\multicolumn{1}{|c|}{Battery Degradation Factor (\%)}     & 3.5    \\ \hline
\multicolumn{1}{|c|}{Battery Life (Years)}                & 10     \\ \hline
\multicolumn{1}{|c|}{2020 Electricity Price (\$/kWh)}     & 0.113 \\ \hline
\multicolumn{1}{|c|}{Analysis Period (Years)}             & 30     \\ \hline
\multicolumn{1}{|c|}{Service Time (Years)}                & 25     \\ \hline
\multicolumn{1}{|c|}{Down Payment (\%)}                   & 10     \\ \hline
\multicolumn{1}{|c|}{General Inflation Rate (\%)}             & 2.5    \\ \hline
\multicolumn{1}{|c|}{Real Discount Rate}                      & 1.95   \\ \hline
\multicolumn{1}{|c|}{Nominal Discount Rate}                      & 4.5    \\ \hline
\multicolumn{1}{|c|}{Average EV Size (kWh)}               & 68.7   \\ \hline
\multicolumn{1}{|c|}{Average EV Range (km)}               & 355    \\ \hline
\multicolumn{1}{|c|}{Efficiency Improvement (\%)}         & 31.7   \\ \hline
\multicolumn{1}{|c|}{Marginal Tax Rate (\%)}              & 20     \\ \hline
\multicolumn{1}{|c|}{Real Electricity Price Escalation Rate (\%)}                    & 0      \\ \hline
\multicolumn{1}{|c|}{Nominal Electricity Price Escalation Rate (\%)}                    & 2.5    \\ \hline
\end{tabular}
\end{table}

\section{Acknowledgement}
This research work is supported by the U.S. Department of Energy's award under grant number DE-EE0008851.

\bibliographystyle{elsarticle-num} 

\end{document}